\documentstyle[12pt]{article}

\def\thefootnote{\fnsymbol{footnote}}
\catcode`\@=11
\def\numberbysection{\@addtoreset{equation}{section}
        \def\theequation{\thesection.\arabic{equation}}}
\def\beq{\begin{equation}}
\def\eeq{\end{equation}}
\def\barr{\begin{eqnarray}}
\def\beqa{\begin{eqnarray}}
\def\earr{\end{eqnarray}}
\def\eeqa{\end{eqnarray}}
\def\winf{W_{1+\infty}\ }
\def\u1{\widehat{U(1)}}
\def\v{V\,}
\def\w{W\,}
\def\vb{{\overline V}\,}
\def\wb{{\overline W}\,}

\def\reps{ representations }
\def\scr{\scriptstyle}

\newcommand{\secn}[1]{Section~\ref{#1}}

\newcommand{\eq}[1]{Eq.~(\ref{#1})}

\newcommand{\nl}{\nonumber \\}

\renewcommand{\theequation}{\thesection.\arabic{equation}}
\newcommand{\EQ}{\begin{equation}}
\newcommand{\EN}{\end{equation}}
\newcommand{\bea}{\begin{eqnarray}}
\newcommand{\ena}{\end{eqnarray}}

\newcommand{\ri}{\rm i}
\newcommand{\re}{\rm e}

\newcommand{\NP}[1]{{\it Nucl.\ Phys.\ }{\bf #1}}
\newcommand{\PL}[1]{{\it Phys.\ Lett.\ }{\bf #1}}

\newcommand{\CMP}[1]{{\it Comm.\ Math.\ Phys.\ }{\bf #1}}
\newcommand{\PR}[1]{{\it Phys.\ Rev.\ }{\bf #1}}
\newcommand{\PRL}[1]{{\it Phys.\ Rev.\ Lett.\ }{\bf #1}}

\newcommand{\IJMP}[1]{{\it Int.\ Jour.\ Mod.\ Phys.\ }{\bf #1}}
\renewcommand{\thefootnote}{\fnsymbol{footnote}}
\numberbysection

\begin{document}
\begin{titlepage}
\begin{center}
\hfill DFTT 07/96 \\
\hfill hep-th/yymmddd \\
\vskip .3 in
{\Large \bf Algebraic bosonization: the study of the \\
Heisenberg and Calogero-Sutherland models}
\vskip 0.8cm
Marialuisa~FRAU,~~Stefano~SCIUTO \\
\vskip 0.3cm
{\em      Dipartimento di Fisica Teorica,
          Universit\`a di Torino,\\
	     and I.N.F.N. Sezione di Torino \\
          Via P. Giuria 1, I-10125 Torino, Italy}
\vskip 0.6cm
Alberto LERDA \\
\vskip 0.3cm
{\em       Dipartimento di Scienze e Tecnologie Avanzate
           \footnote{II Facolt\`a di Scienze
            M.F.N., Universit\`a di Torino (sede di Alessandria),
Italy}
	   and \\
           Dipartimento di Fisica Teorica, Universit\`a di Torino,\\
           and I.N.F.N., Sezione di Torino \\
           Via P.Giuria 1, I-10125 Torino, Italy}
\vskip 0.6cm
Guillermo~R.~ZEMBA \\
\vskip 0.3cm
{\em       Centro At\'omico Bariloche \\
           8400 - San Carlos de Bariloche ~(R\'{\i}o Negro),
Argentina}

\end{center}
\vskip 1cm
\begin{abstract}
\noindent
We propose an approach to treat $(1+1)$--dimensional fermionic
systems based on the idea of {\it algebraic bosonization}. 
This amounts to decompose the elementary low-lying excitations 
around the Fermi surface in terms of basic building blocks 
which carry a representation of the 
$\winf \times {\overline \winf}$algebra, which is the 
dynamical symmetry of the Fermi quantum incompressible fluid.
This symmetry simply expresses the local particle-number
current conservation at the Fermi surface.
The general approach is illustrated in detail in two examples: the
Heisenberg and Calogero-Sutherland models, which allow for 
a comparison with the exact Bethe Ansatz solution.
\end{abstract}
\vfill
\hfill March 1996
\end{titlepage}
\pagenumbering{arabic}
\renewcommand{\thefootnote}{\arabic{footnote}}
\setcounter{footnote}{0}
\setcounter{page}{1}

\section{Introduction}
\bigskip

There are many $(1+1)$--dimensional models of non-relativistic
fermions which are of contemporary interest, either for theoretical
reasons, or because of their applicability to specific
condensed matter or statistical systems (for modern introductions to
this subject see for example Refs.~\cite{frad,fermi}).
Some of these models admit an exact solution by
the application of the Bethe Ansatz technique \cite{bethe}
(for a general review see, for example, Ref.~\cite{kore}).
Although this method is very powerful and provides a deep conceptual
insight that most approximations miss, in some cases it may be
difficult to extract explicit results from it. Moreover, there are many
interesting fermionic systems that cannot be exactly solved by the
Bethe Ansatz.

In this paper, we present a method that can partially circumvent these
problems leading to simple and tractable expressions, at least in
perturbation theory, and which can be also applied to non-integrable
models. We shall call this procedure {\it algebraic bosonization},
which can be described as a sequence of simple steps. The first
is to identify the Fermi surface (the set of left
and right Fermi points in the simplest case)
of the $(1+1)$--dimensional systems under consideration, and then
study its small fluctuations \cite{hald1,froh,boson}, 
{\it i.e.} those many-body configurations of low momenta in 
the vicinity of the Fermi surface.
The dynamics of such configurations is governed
by an effective hamiltonian ${\cal H}$
which is, in general, simpler than the original one.
In the thermodynamic limit of a large number of fermions $N$
at constant density $\rho$, ${\cal H}$ scales as
a power series in $1/N$. As a consequence, only the first
few terms of this expansion must be considered to achieve
a given degree of accuracy.
In particular, for a gapless system, {\it i.e.} with a linear
dispersion relation around the Fermi points, the $1/N$ term
of the effective hamiltonian identifies a conformal field theory
\cite{bpz}. In fact, one has
\beq
{\cal H}_{(1/N)} = \frac{2\pi}{N}\,\rho\,v\,
\left(L_0+{\overline L}_0\right)
\label{confo}
\eeq
where $v$ is the Fermi velocity, and $L_0$ and ${\overline L}_0$
are the zero modes of the right and left Virasoro (conformal)
algebras.
The validity of \eq{confo} has been established for the general
class of the so-called
Luttinger systems \cite{hald1,fermi}, and for all gapless models
solvable by the Bethe Ansatz \cite{hald2,kofr,kaya}.

The spectrum of ${\cal H}_{(1/N)}$ follows directly from the
representation theory of the Virasoro algebra, which
appears as the dynamical symmetry (or spectrum generating algebra)
of the effective hamiltonian to order $1/N$.
At this point, it is natural to ask whether a dynamical
symmetry exists even if the subleading $O(1/N^2)$-terms of
${\cal H}$ are taken into account.
Given that these terms originate partly from the interactions among
the fermions, and partly from the non-linearity of the dispersion
curve around the Fermi surface, the Virasoro algebra is insufficient
to describe this new situation and some extension of it becomes
necessary.

To get a hint of what the new algebra might be, one can note the
following observations. In the
thermodynamic limit, the dynamics of the zero-dimensional Fermi
surface becomes semiclassical \cite{hald1,froh,boson} 
and the one-dimensional Fermi sea behaves as a droplet of
an incompressible classical fluid in momentum space.
Obviously, this can be thought also as a
one-dimensional section of a two-dimensional incompressible
droplet. The classical configurations of the latter are
characterized by the dynamical symmetry under the
area-preserving diffeomorphisms, which
generate the so-called $w_\infty$ algebra \cite{shen}.
In view of these considerations, we are led to propose as
dynamical symmetry of the effective hamiltonian the
$\winf$algebra \cite{shen,kac1}, which is a quantum version of the
$w_\infty$ algebra generated by the small fluctuations of the Fermi
surface of the $(1+1)$-dimensional system.
For the systems we shall consider, $\winf$ simply expresses the
local conservation of the particle-number current at each Fermi
point.

This situation resembles the physics of the quantum Hall effect (for
a review see, {\it e.g.} Ref.~\cite{prange}),
where the role of the Fermi sea (in configuration space)
is taken by the Laughlin's quantum incompressible fluid \cite{laugh}.
Indeed, the latter has been shown to possess the $\winf$dynamical
symmetry generated by the edge excitations \cite{ctz3}, which
can be bosonized using the $\winf$symmetry \cite{ctz4}.
The $\winf$algebra is a linear and infinite dimensional
extension of the Virasoro algebra,
containing generators $\v^i_n$ (with $n \in{\bf Z}$ and $i \geq 0$)
of arbitrary integer conformal spin $i+1$ (in this notation,
the standard Virasoro generators $L_n$ are denoted by $\v_n^1$).

The second step of our procedure is to show that the complete
effective hamiltonian ${\cal H}$ of the fermionic system
displays a $\winf$structure. This amounts to prove that the
subleading
part in the $1/N$ expansion can be written entirely in terms of
$\v^i$ currents.
If this condition is met, then the Hilbert space of the effective
hamiltonian is described by a set
of unitary, irreducible, highest-weight representations of
the $\winf$algebra, which are known and completely classified
\cite{kac1}. Hence the spectrum of the low-lying excitations
can be readily obtained.

We would like to stress that the purpose of this procedure is not to
simply
rewrite the effective hamiltonian in a different
fashion, but instead to extend to all orders in $1/N$ the abelian
bosonization of the Luttinger models
\cite{lutt,matlib,lupe,hald1,fermi}.
In fact, if the fundamental degrees of freedom of the effective
theory are the modes of the currents $V^i$, then any realization of
these can be chosen for convenience. In particular, a realization of the
$\winf$algebra in terms of a bosonic field can then be used
to describe the system in place of the original fermionic degrees of
freedom.
In this way, one can take advantage of the existence a free
parameter,
the compactification radius of the bosonic field,
to diagonalize the {\it entire} effective hamiltonian.

In this paper we illustrate in detail this procedure in two different
systems: the Heisenberg and the Calogero-Sutherland models.
The former can be mapped by means of a Jordan-Wigner transformation
into a theory of fermions on a lattice with a short-range
interaction;
the latter is, instead, a continuum theory
with long-range interactions.
Despite this difference at the microscopic level,
the effective hamiltonians ${\cal H}$ of the two models turn out to
have
the same structure, namely that of an interacting Luttinger
liquid \cite{hald1,fermi}.
By exposing the $\winf$structure of ${\cal H}$, we will
be able to use the $\winf$representation theory to
find the spectrum of their low-energy excitations.

The paper is organized as follows. In \secn{linear} we
derive the effective theory of the Heisenberg and
Calogero-Sutherland models, using
spinless fermions as microscopic degrees of freedom.
In \secn{w} we give a brief summary of the main properties of the
$\winf$algebra, and present explicitly both its fermionic and bosonic
realizations which will be useful in the sequel; some technical
and mathematical details on the $\winf$algebra and its
representations
are, instead, collected in the Appendix. In \secn{csw} we study
the $\winf$structure of
the Calogero-Sutherland model, and compare our results with those
obtained
from the exact Bethe Ansatz solution. We also comment on
the connections between the two methods.
In \secn{heisw} we apply the same procedure to the Heisenberg model,
up to
order $1/N^2$ in the presence of an external magnetic field $B$, and
up to order $1/N^3$ when $B=0$. Finally, in \secn{concl} we present
our conclusions.

\vskip 1.5cm
\section{The effective theory of the Heisenberg
and the Calogero-Sutherland models}
\label{linear}
\bigskip

In this section we will derive the effective
hamiltonian for the Heisenberg and the Calogero-Sutherland
models in $(1+1)$-dimensions, and show that both of them have
the structure of an interacting Luttinger liquid \cite{hald1,fermi}.
This result is achieved by using spinless fermions as microscopic
degrees of freedom, and linearizing their
dispersion relations around the Fermi points. We begin by discussing
in detail this procedure for the Heisenberg model, leaving the
analysis of the Calogero-Sutherland model for later.

\bigskip
\noindent{\bf The Heisenberg model}

The Heisenberg model describes the exchange interactions among
spins localized on the sites of a lattice, with hamiltonian
\beq
H = \sum_{<j,k>} J~{\bf S}_j\cdot {\bf S}_k = \sum_{<j,k>} \left[
\frac{J}{2}\left(S_j^+\,S_k^- + S_j^-\,S_k^+\right) +J\,S_j^z\,S_k^z
\right]~~~.
\label{heis}
\eeq
Here $J$ is a coupling constant and the symbol $<j,k>$ denotes, as
usual, a pair of nearest-neighbor sites. The spin operators
${\bf S}_j\equiv\{S_j^{+},
S_j^{-},S_j^{z}\}$
close the  $SU(2)$ algebra
\bea
\left[ S_j^+ \, , \, S_k^- \right ]
&=& 2\, S_j^z\,\delta_{j,k} ~~~,\nl
 \left[\, S_j^\pm\, , \, S_k^z\,\right]
&=& \pm\, S_j^\pm\,\delta_{j,k}~~~,
\label{su2}
\ena
and belong to the spin $s$ representation, {\it i.e.}
${\bf S}_j\cdot {\bf S}_j=s(s+1)$ for all $j$.
In the following we will consider the case $s=1/2$ only.
Of particular interest
is the deformation of the model (\ref{heis})
known as the XXZ model
\cite{YangYang}. This is characterized by a spin
anisotropy in the $z$ direction encoded in the difference
between the values of the coupling
constants of the two terms of the square brackets of \eq{heis}.

For a one dimensional chain with $N$ sites,
the hamiltonian of the XXZ model
in an external magnetic field $B$ is
\beq
H= H_{XX}+ H_Z + H_B
\label{heisham0}
\eeq
with
\bea
H_{XX}&=&\frac{J}{2}\sum_{j=1}^N\left(S_j^+\,S_{j+1}^-+S_j^-\,
S_{j+1}^ +\right) ~~~,\nl
H_{Z}&=&J_z\sum_{j=1}^N S_j^z\,S_{j+1}^z
{}~~~,\label{heisham} \\
H_{B}&=& - B\sum_{j=1}^N S_j^z ~~~,\nonumber
\ena
where $J_z \,(\not= J)$ is the coupling constant that gives
the anisotropy in the $z$ direction.
We will assume periodic boundary conditions
on the chain, {\it i.e.}
${\bf S}_{N+1}\equiv{\bf S}_1$, and set $J=-1$ in $H_{XX}$
for convenience. The model (\ref{heisham0})
can be exactly solved by the Bethe Ansatz for general 
values of the coupling constant $J_z$ \cite{YangYang},
and a large amount of physical information can
be obtained from this solution \cite{kore}.
In particular one finds that the system is gapless in the
antiferromagnetic regime ($J_z>-1$) with low-lying excitations
above the ground state described by a two-dimensional
conformal field theory. This theory captures the universal
physical behavior of the system at large distances, such as
the critical exponents, and hence it must be regarded as the
effective theory corresponding to the original microscopic model.
In what follows, we will study
this conformal field theory and its $\winf$algebra extension.
To do so, we will follow essentially the same strategy
outlined in the seminal paper by Luther and Peschel \cite{lupe}:
we will first transform the
Heisenberg model (\ref{heisham}) into a system of interacting
spinless fermions with a linear dispersion relation near the
Fermi surface, and then bosonize it.

The first step is
easily achieved: we can obtain a fermionic description
of the spin-1/2 Heisenberg model by means of the
Jordan-Wigner transformation.
In fact, if we introduce spinless fermionic oscillators, ${\hat
\psi}_j$,
with standard anticommutation relations
\bea
\{{\hat \psi}_j\, ,\, {\hat \psi}_k\} &=&
\{{\hat \psi}_j^\dagger\, ,\, {\hat \psi}_k^\dagger\}
=0 ~~~,\nl
\{{\hat \psi}_j\, ,\, {\hat \psi}_k^\dagger\} &=& \delta_{j,k}~~~,
\label{psipsi}
\ena
then, the algebra (\ref{su2}) can be identically satisfied
by defining
\bea
S_j^+&=&\exp\left({\rm i}\pi\sum_{k=1}^{j-1}n_k\right)~{\hat
\psi}^\dagger_j
{}~~~,\nl
S_j^-&=&\exp\left(-{\rm i}\pi\sum_{k=1}^{j-1}n_k\right)~{\hat \psi}_j
{}~~~,\label{jorwig}\\
S_j^z&=&{\hat \psi}_j^\dagger\,{\hat \psi}_j -\frac{1}{2}
{}~~~,\nonumber
\ena
where $n_j\equiv{\hat \psi}^\dagger_j\,{\hat \psi}_j$
is the fermion number at site $j$.
The operators
$\exp\left(\pm{\rm i}\pi\sum\limits_{k=1}^{j-1}n_k\right)$
are cocycle factors that must be introduced to correct the fermionic
statistics of ${\hat \psi}_j$ and ${\hat \psi}_j^\dagger$, in such a
way
that $S_j^{\pm}$ at different sites commute with each other,
as required by \eq{su2}.

To properly write the hamiltonian $H$ in the
fermionic representation, we observe that the
expectation value of $S_j^z$ in the antiferromagnetic
ground state, $\langle S_j^z \rangle$,
is not vanishing in the presence of a magnetic field.
Thus, in general we can write
\beq
S_j^z= ~:{\hat \psi}_j^\dagger\,{\hat \psi}_j : +
\frac{\sigma}{2}~~~,
\label{sigma}
\eeq
where colons denote the normal ordering with respect to the ground
state, and $\sigma$ the magnetization per site.

Using this definition and \eq{jorwig}, we see that
the three terms of \eq{heisham} become
\bea
H_{XX}&=&
-\ \frac{1}{2} \sum_{j=1}^N\left(:{\hat \psi}_j^\dagger\,{\hat
\psi}_{j+1}
:+:{\hat \psi}_{j+1}^\dagger\,{\hat \psi}_j :\right)~~~,
\label{hxy} \\
H_Z&=& J_z\sum_{j=1}^N \left(:{\hat \psi}_j^\dagger\,{\hat \psi}_j:~
:{\hat \psi}_{j+1}^\dagger\,{\hat \psi}_{j+1} :\right)
+J_z\,\sigma\sum_{j=1}^N :{\hat \psi}_j^\dagger\,{\hat \psi}_j :~~~,
\label{hz} \\
H_B &=& - B \sum_{j=1}^N:{\hat \psi}_j^\dagger\,{\hat \psi}_j :~~~,
\label{hb}
\ena
where we have also normal-ordered $H_{XX}$ and dropped
irrelevant (finite) constants
\footnote{Notice that the introduction of the normal ordering
in $H_{XX}$ yields a finite
constant which is the ground state energy of the $XX$
model.}.
The total hamiltonian $H=H_{XX}+H_Z+H_B$ describes a system of
interacting spinless fermions, and consists of terms that
are quadratic and quartic in the ${\hat \psi}$'s . Thus, it
is useful to distinguish between them and write
\beq
H= H_0 +H_I~~~,
\label{h0hi}
\eeq
where $H_0$ denotes the quadratic part and $H_I$ the four-fermion
interaction. Our attitude will be to focus first on $H_0$ and then
treat $H_I$ as a perturbation. Notice, however, that the ``free''
hamiltonian $H_0$ actually depends both on the coupling constant
$J_z$ and on the magnetic field $B$,
since both $H_Z$ and $H_B$ contribute to $H_0$.

{}From now on, we will assume for simplicity that the chain
has an even number of sites and a unit lattice spacing.
Thus, if $L$ denotes the total length of the chain, we have
$$
L=N=2\,M~~~,
$$
with $M$ being a positive integer number.
The Fourier transform of the fermion ${\hat \psi}_j$ is then given by
\beq
{\hat \psi}_j=\frac{1}{\sqrt{N}}\sum_{n}
\psi_n\,{\re}^{{\ri}\,k_n\,j}~~~,
\label{fourier}
\eeq
where $k_n\equiv(2\pi n/N)$ is the momentum of the mode $\psi_n$.
To ensure the appropriate boundary conditions,
the index $n$ must be integer if $M$ is odd, and half-integer if $M$
is even. Furthermore, due to the periodicity of the lattice,
the sum over $n$ in \eq{fourier} must be restricted
to the first Brillouin zone. In the following, without any loss of
generality, we will consider only the case $M$ odd.

Inserting \eq{fourier} into Eqs. (\ref{hxy})-(\ref{hb}) and
decomposing
the result according to \eq{h0hi}, we get
\bea
H_0 &=& -\sum_{n=-M+1}^M
\big[f(n)+(B-J_z\,\sigma)\big]
\,:\psi_n^\dagger\,\psi_n:~~~,
\label{h0} \\
H_I
&=&\frac{J_z}{N}\sum_{n,n',m,m'=-M+1}^M
f(n-n')
\,:\psi_{n'}^\dagger\,\psi_n:~:\psi_{m'}^\dagger\,\psi_m:
\,\delta(n'-n+m'-m)
\nl
&&+\ \frac{J_z}{N}
\sum_{n,n',m,m'=-M+1}^M
{\re}^{{\ri}\frac{2\pi}{N}(n'-n)}\,
:\psi_{n'}^\dagger\,\psi_n:~:\psi_{m'}^\dagger\,\psi_m:
\label{hi} \\
&&\times \ \Big[\delta(n'-n+m'-m-N)+
\delta(n'-n+m'-m+N)\Big]~~~,
\nonumber
\ena
where
\beq
f(n)=\cos\left(\frac{2\pi}{N}\,n\right)~~~,
\label{fn}
\eeq
and $\delta(n)$ stands for the Kronecker delta $\delta_{n,0}$.
The last two lines of \eq{hi} are the Umklapp terms, which are
characterized by the fact that the four momenta of the fermions add
up to $\pm2\pi$, and thus satisfy momentum conservation modulo a
reciprocal lattice vector.

When $B=0$ (and hence $\sigma=0$), $H_0$ simply describes
a free fermionic system with ground state given by
$$
|\Omega \rangle_0 =\psi_{-n_F^0}^\dagger \cdots \psi_{n_F^0}^\dagger
|0\rangle~~~,
$$
where $|0\rangle$ is the Fock vacuum of the fermionic oscillators,
and $n_{F}^0=(M-1)/2$ is the Fermi point.
The introduction of a (not too strong)
magnetic field $B$ does not destroy this
structure, merely shifting the Fermi level, $n_F^0\rightarrow
n_F$. More precisely, using \eq{sigma} one finds that
\beq
n_F=\frac{M}{2}(1+\sigma)-\frac{1}{2}~~~,
\label{nf}
\eeq
so that the ground state of $H_0$ in the presence of a
magnetic field is
\beq
|\Omega \rangle =\psi_{-n_F}^\dagger \cdots \psi_{n_F}^\dagger
|0\rangle~~~.
\label{ground}
\eeq
The quantity $n_F/(N/2)$ is usually called the filling factor
\cite{frad},
which in the absence of an external magnetic field
takes the value $1/2$ in the thermodynamic limit.
The two isolated points $\pm n_F$ form the Fermi surface of the
system.
For convenience here and in the following, we assume that the
quantity $(M\sigma/2)$ is an
integer in such a way that $n_F$ is simply obtained from
$n_F^0$ with an integer shift \footnote{Notice that this requirement
implies that $\sigma$ has to be quantized in units of $2/M=1/N$, but
the effects of this discretization actually
disappear in the thermodynamic limit $N\to\infty$.}.
Later on, we will determine the precise relation between $\sigma$ and
the magnetic field $B$ (see \eq{bsigma}), but for the time being this
relation is not necessary.
We only notice here that the shift of the Fermi surface induced
by the magnetic field guarantees
that the Umklapp terms do not contribute to the low-energy effective
hamiltonian, as we will see momentarily.

In general, only the oscillators near the Fermi points $\pm n_F$
play an important role in physical processes. In fact, they
produce the low-energy excitations above the ground state
$|\Omega\rangle$ and determine the large-distance properties
of the system which are described by the
effective theory. In order to write the hamiltonian
for this effective theory, we define
shifted fermionic operators associated to the small fluctuations
around each Fermi point according to
\bea
a_r\equiv\psi_{n_F+r}~~~&,&~~~
a_r^\dagger\equiv\psi_{n_F+r}^\dagger
{}~~~,\nl
b_r\equiv\psi_{-n_F-r}~~~&,&~~~
b_r^\dagger\equiv\psi_{-n_F-r}^\dagger~~~.
\label{arbr}
\ena
The quantity $2\pi\left(r-1/2\right)/N$ ($-2\pi\left(r-1/2\right)/N$)
represents the momentum of the oscillator $a_r$ ($b_r$) relative to
the
right (left) Fermi point.
The integer index $r$ can only vary in a
finite range, say between $-\Lambda_0$ and $+\Lambda_0$
where $\Lambda_0$ is a bandwidth cut-off. We choose it such
that
$\Lambda_0 << n_F$, and $\Lambda_0=o(n_F)=o(N)$ in the thermodynamic
limit
$N\to\infty$.
Roughly speaking,
$\Lambda_0$ indicates how far from the Fermi points one can go
without leaving the effective regime. Clearly the oscillators $a$ and
$b$ in \eq{arbr}
form two independent sets and define two independent and finite
branches of excitations, one around the left and one around
the right Fermi points. They are such that
\bea
a_r|\Omega\rangle = 0~~~&,&~~~
b_r|\Omega\rangle = 0~~~~{\rm for}~~r=1,2,\ldots,\Lambda_0~~~,
\nl
a_s^\dagger|\Omega\rangle = 0~~~&,&~~~
b_s^\dagger|\Omega\rangle = 0 ~~~~{\rm for}~~
s=0,-1,-2,\ldots,-\Lambda_0~~~.
\label{abvac1}
\ena

The procedure to find the effective hamiltonian ${\cal H}$
corresponding to $H$ is now simple. The first step is to select
all terms in $H$ containing oscillators whose
index lies in the range $[-\Lambda_0,+\Lambda_0]$
around each Fermi point.
For example, for the quadratic part of the hamiltonian, $H_0$, we
get
\beq
{\cal H}_0 = -\sum_{r=-\Lambda_0}^{\Lambda_0} \Big[
f(r+n_F)+(B-J_z\,\sigma)\Big]\left(
:a^\dagger_r\,a_r:+:b^\dagger_r\,b_r:\right)~~~,
\label{ho}
\eeq
where the normal ordering is defined with respect to $|\Omega\rangle$
according to \eq{abvac1}.
After using Eqs. (\ref{fn}) and (\ref{nf}), we expand
the right hand side of \eq{ho} in powers of $1/N$ to obtain
\bea
{\cal H}_0 &=& \left(-B+J_z\,\sigma+
\sin\frac{\pi\sigma}{2}\right)
\sum_{r=-\Lambda_0}^{\Lambda_0}
\left(:a^\dagger_r\,a_r:+:b^\dagger_r\,b_r:\right)\nl
&&+\ \frac{2\pi}{N}\,\cos\frac{\pi\sigma}{2}
\sum_{r=-\Lambda_0}^{\Lambda_0}\left(r-\frac{1}{2}\right)
\left(:a^\dagger_r\,a_r:+:b^\dagger_r\,b_r:\right)\label{h0+} \\
&&-\ \frac{1}{2}\left(\frac{2\pi}{N}\right)^2\,
\sin\frac{\pi\sigma}{2}
\sum_{r=-\Lambda_0}^{\Lambda_0}\left(r-\frac{1}{2}\right)^2
\left(:a^\dagger_r\,a_r:+:b^\dagger_r\,b_r:\right)\nl
&&-\ \frac{1}{6}\left(\frac{2\pi}{N}\right)^3\,
\cos\frac{\pi\sigma}{2}
\sum_{r=-\Lambda_0}^{\Lambda_0}\left(r-\frac{1}{2}\right)^3
\left(:a^\dagger_r\,a_r:+:b^\dagger_r\,b_r:\right)\nl
&&+\ O\left(\frac{1}{N^4}\right)~~~.
\nonumber
\ena

We remark that this expansion is meaningful
because the sum over $r$ has a finite range and $N$
is assumed to be large. The thermodynamic limit is correctly
defined because, according to our assumptions,
$\Lambda_0=o(N)$ when $N\to \infty$. In particular, we stress that
the $1/N$-term of \eq{h0+} has a linear
dispersion relation, signaling the fact
that the system is gapless. The $O(1/N^2)$-terms of this expansion
are higher order corrections and will be analyzed
in the following sections using the structure of the $\winf$algebra.

Let us now turn to the interaction term, \eq{hi}.
To write the corresponding effective hamiltonian ${\cal H}_I$,
we distinguish among three cases:
\begin{enumerate}
\item when the momentum exchanged in the interaction is
small, that is $|n'-n|\sim 0$;
\item when the exchanged momentum is roughly
twice the Fermi momentum, that is $|n'-n|\sim 2n_F$;
\item when Umklapp processes take place.
\end{enumerate}

The first case corresponds to a
forward scattering, whilst the second to a backward scattering;
both kinds of processes are described by
the first line of \eq{hi}. The Umklapp terms instead,
originate only from
the last two lines of \eq{hi}. According to this
classification, it is natural
to decompose the effective hamiltonian as follows
$$
{\cal H}_I = {\cal H}_{forw}+{\cal H}_{back}+{\cal H}_{Umkl}~~~.
$$

Let us first consider the forward scattering terms.
These arise from \eq{hi} with the following choice of
configuration of fermionic indices
\bea
n=\pm\left(n_F+r\right)~~&,&~~n'=n-\ell~~~,\nl
m=\pm\left(n_F+s\right)~~&,&~~m'=m+\ell~~~,
\label{confforw}
\ena
with
\bea
|r|\leq \Lambda_0 ~~~&,&~~~ |s|\leq \Lambda_0 ~~~,\nl
|r\mp\ell|\leq \Lambda_0~~~&,&~~~|s\pm\ell|\leq \Lambda_0~~~.
\label{ranges}
\ena
In \eq{confforw} we can take either sign independently,
so that we find
\bea
{\cal H}_{forw} &=&
\frac{J_z}{N}\,{\sum_\ell} '
\sum_{r,s=-\Lambda_0}^{\Lambda_0}f(\ell)
\left[:a^\dagger_{r-\ell}\,a_r:\,:a^\dagger_{s+\ell}\,a_s:
+:b^\dagger_{r+\ell}\,b_r:\,:b^\dagger_{s-\ell}\,b_s: \right.
\nl
&&+\ \left.:a^\dagger_{r-\ell}\,a_r:\,:b^\dagger_{s-\ell}\,b_s:
+:b^\dagger_{r+\ell}\,b_r:\,:a^\dagger_{s+\ell}\,a_s:
\right]~~~.
\label{hifor}
\ena
Here the symbol $\,\sum'$ means that
the sum over $\ell$ is restricted to those values that
satisfy the constraints (\ref{ranges}).
Of the four terms appearing in the square bracket of \eq{hifor},
the first two represent the forward scattering among
four particles belonging
to the same branch of the dispersion curve (right or left),
while the second two represent the forward scattering between
two pairs of particles belonging to different branches.

Next, we examine the backward scattering part of the effective
hamiltonian. It corresponds to
the following configurations of indices for the fermions in \eq{hi}
\bea
n=\pm\left(n_F+r\right)~~&,&~~n'=n \mp 2n_F-\ell~~~,\nl
m=\mp\left(n_F+s\right)~~&,&~~m'=m\pm 2n_F+\ell~~~,
\label{confback}
\ena
where, once again, $r$, $s$ and $\ell$ satisfy \eq{ranges}.
To fulfill momentum conservation,
we must take in \eq{confback} either all upper or all lower signs,
thus generating two different structures
in the effective hamiltonian.
In fact, we find
\bea
{\cal H}_{back} &=&
\frac{J_z}{N}\,{\sum_\ell}'
\sum_{r,s=-\Lambda_0}^{\Lambda_0} \left[
f\left(2n_F+\ell\right)\,
b^\dagger_{-r+\ell}\,a_r\,a^\dagger_{-s+\ell}\,b_s \right. \nl
&&+\ \left. f\left(-2n_F+\ell\right)\,
a^\dagger_{-r-\ell}\,b_r\,b^\dagger_{-s-\ell}\,a_s \right]~~~.
\label{hibac}
\ena

Finally, we consider the Umklapp terms, which, as mentioned above,
originate from the last two lines of \eq{hi}.
These terms are important only if the band is half-filled,
in which case all four fermions can be near the
Fermi surface. If the band is not half-filled, the Umklapp processes
do not contribute to the effective hamiltonian.
This is precisely what happens for the Heisenberg
model in the presence of a magnetic field $B$. In fact, according to
\eq{nf} for $\sigma\not=0$, the difference
$(4n_F-N)$ is of order $N$ in the thermodynamic limit.
Therefore, the delta functions
in the last line of \eq{hi} cannot have vanishing argument if $n'$,
$n$, $m'$ and $m$ differ from $\pm n_F$ by a quantity
$|\ell|\leq 2\Lambda_0=o(N)$ for $N\to \infty$. This statement is
true for any non-zero magnetic field $B$, even if very small.
Thus, in the following, we will always neglect the Umklapp
contributions,
taking
$$
{\cal H}_{Umkl} = 0 ~~~.
$$
Notice that the
case with no magnetic field also shares this property
provided it is defined as the limit $B\to 0$
of the case with non-vanishing $B$.

The next step to find the low-energy theory
is the crucial one: we remove the bandwidth cut-off
$\Lambda_0$ by sending it to infinity, as in the mapping of
the Tomonaga model \cite{tomo}
into the Luttinger model \cite{lutt}. However,
in order to avoid the introduction of spurious low-energy states
(the $O(1/N^2)$-corrections do bend the dispersion curve),
we will keep always $\Lambda_0<<N$.
For ease of notation, when $\Lambda_0$ and $N \to \infty$,
we will write the free effective hamiltonian
${\cal H}_0$ simply as in \eq{h0+}
with all sums extended from $-\infty$ to $+\infty$ but with
the limit $N\to\infty$ left implicit.
Even though ${\cal H}_0$ now contains
new oscillators, it should be clear that it still acts
on low-energy states only,
with particle and hole momenta bounded by $\Lambda_0<<N$.
Nonetheless, the extension of the dispersion curve to
infinity is not free of consequences. In fact,
since now Eqs. (\ref{abvac1}) hold for {\it any} integer
$r$ and $s$, the ground state $|\Omega\rangle$ corresponds
to the surface of two {\it infinite}
Fermi seas (one left and one right). This fact will
play a crucial role when describing the algebraic properties
of the effective theory, leading to an easier and more
elegant mathematical interpretation,
as we shall see momentarily.

When $\Lambda_0\to\infty$, it is natural to define the following
two chiral Weyl fermions
\beq
F_+(\theta) = \sqrt{\frac{2\pi}{N}}\sum_{r=-\infty}^\infty
a_r~{\re}^{{\ri}\left(r-\frac{1}{2}\right)\theta}~~~,
\label{f+}
\eeq
and
\beq
F_-(\theta) = \sqrt{\frac{2\pi}{N}}\sum_{r=-\infty}^\infty
b_r~{\re}^{{\ri}\left(r-\frac{1}{2}\right)\theta}~~~.
\label{f-}
\eeq
Since the index $r$ is integer, these fields
satisfy antiperiodic boundary conditions on a
circle of radius $R=N/2\pi$, parametrized by the angle $\theta\in
[0,2\pi)$. Using $F_+$ and $F_-$, it is straightforward
to check that the $1/N$-term
in ${\cal H}_0$ can be written as
\beq
{\cal H}_0\,\Big|_{1/N} = \frac{1}{2\pi}\int_0^{2\pi}d\theta~
:\,\left[F_+^\dagger\,(-{\ri}\partial_\theta)F_+
+ F_-^\dagger\,(-{\ri}\partial_\theta)F_-\right]\,:~~~.
\label{dirac}
\eeq
This is the Dirac hamiltonian for a free
relativistic fermion in $(1+1)$-dimensions given by
$$
\Phi=\left(\matrix{
 F_+ \cr
 F_- \cr}\right)~~~.
$$
The other terms of ${\cal H}_0$ can also be nicely written
in terms of this Dirac field.
As it is well-known, in $(1+1)$-dimensions a Dirac fermion is
equivalent to a scalar boson through the abelian bosonization
procedure.
Thus, the free hamiltonian ${\cal H}_0$
can be given an equivalent
description using bosons instead of fermions.
Actually, we shall see that this is possible
even for the interaction hamiltonian ${\cal H}_I$.

To do so, we must
analyze ${\cal H}_{forw}$ and ${\cal H}_{back}$
when the bandwidth is extended to infinity.
For the forward scattering part, \eq{hifor} simply leads to
\bea
{\cal H}_{forw} &=&
\frac{J_z}{N}\,
\sum_{\ell,r,s=-\infty}^{\infty}
f(\ell)
\left[:a^\dagger_{r-\ell}\,a_r:\,:a^\dagger_{s+\ell}\,a_s:
+ :b^\dagger_{r+\ell}\,b_r:\,:b^\dagger_{s-\ell}\,b_s: \right.
\nl
&&+\ \left. 2\,:a^\dagger_{r-\ell}\,a_r:\,:b^\dagger_{s-\ell}\,b_s:
\right]~~~,
\label{hifor4'}
\ena
where we have used the property $f(\ell)=f(-\ell)$.
Since all terms in the r.h.s. of \eq{hifor4'}
are already normal ordered in each pair of fermions,
this equation is a good starting point to implement the bosonization
procedure, as we shall discuss in the following sections.

The backward scattering terms, \eq{hibac},
require more attention instead. In fact, before
one can bosonize them, it is necessary to rearrange the $a$ and $b$
oscillators to reconstruct normal ordered pairs.
Sending $\Lambda_0\to\infty$, using the property
$f(\ell\pm N/2)=-f(\ell)$ and
then relabeling the summation index $\ell$,
we can rewrite ${\cal H}_{back}$ as follows
\beq
{\cal H}_{back} =
-\frac{J_z}{N}\,\sum_{\ell,r,s=-\infty}^{\infty}
f(\ell)\left[
b^\dagger_{-r+\ell+\gamma}\,a_r\,a^\dagger_{-s+\ell+\gamma}\,b_s
+a^\dagger_{-r-\ell+\gamma}\,b_r\,b^\dagger_{-s-\ell+\gamma}\,a_s
\right]~~~,
\label{hibac1}
\eeq
where the integer $\gamma$ is defined by
\beq
\gamma=1-\frac{N\sigma}{2}=\frac{N}{2}-2n_F~~~.
\label{gamma}
\eeq
In the square brackets of \eq{hibac1} we have collected together
two terms that
describe backward scattering processes involving
the same exchanged momentum; in fact, according to \eq{gamma},
$$
k=\frac{2\pi}{N}\,(2n_F+\ell+\gamma)~~~,
$$
which is the momentum exchanged by the first term, and
$$
k'=\frac{2\pi}{N}\,(-2n_F+\ell-\gamma)~~~,
$$
which is the momentum exchanged by the second term,
differ by $2\pi$.
Guided by this observation, we normal order ${\cal H}_{back}$
without breaking the square bracket of \eq{hibac1}; this assures
that no divergences appear. Indeed,
after performing standard manipulations
and dropping irrelevant additive constants, we find
$$
{\cal H}_{back} = {\cal H}_{back}^{(4)} + {\cal H}_{back}^{(2)}
$$
where the four- and two-fermion parts are given, respectively, by
\beq
{\cal H}_{back}^{(4)} =
2\,\frac{J_z}{N}\,\sum_{\ell,r,s=-\infty}^{\infty}
f(\ell)
\,:a^\dagger_{-s+\ell+\gamma}\,a_r:\,:b^\dagger_{-r+\ell+\gamma}
\,b_s:~~~,
\label{hibac4}
\eeq
and
\beq
{\cal H}_{back}^{(2)} =
\frac{J_z}{N}\,\sum_{\ell=-\infty}^{\infty}
f\left(\ell\right)\,\left[\left(
\sum_{r=|\ell|+\gamma}^{\infty}
-\sum_{r=-\infty}^{-|\ell|+\gamma-1}\right)\,
\left(:a^\dagger_r\,a_r:+:b^\dagger_r\,b_r:\right)\right]~~~.
\label{hibac2}
\eeq

Since the oscillators are already normal ordered, it is safe to
shift their indices, and also to exchange the order of the sums.
For example, if in \eq{hibac4} we let
$\ell\rightarrow r+s-\ell-\gamma$, we get
\bea
{\cal H}_{back}^{(4)} &=&
2\,\frac{J_z}{N}\,\sum_{\ell,r,s=-\infty}^{\infty}
f(r+s-\ell-\gamma)\,
:a^\dagger_{r-\ell}\,a_r:\,:b^\dagger_{s-\ell}\,b_s: \nl
&=&
2\,\frac{J_z}{N}\,\sum_{\ell,r,s=-\infty}^{\infty}
\left[\cos\pi\sigma\,\cos\frac{2\pi}{N}\left(
r+s-\ell-1\right)\right.
\label{hibac4'}\\
&&-\ \left.
\sin\pi\sigma\,\sin\frac{2\pi}{N}\left(
r+s-\ell-1\right)\right]
:a^\dagger_{r-\ell}\,a_r:\,:b^\dagger_{s-\ell}\,b_s: ~~~,
\nonumber
\ena
where in the last step we have
used Eqs. (\ref{fn}) and (\ref{gamma}).
These manipulations have rendered the operator structure of
${\cal H}_{back}^{(4)}$ identical to that of the last term
of ${\cal H}_{forw}$ in \eq{hifor4'}, the only difference
remaining in the numerical function in front.

It is convenient to simplify also the two-fermion part of
the backscattering hamiltonian. To do so, we first exchange the sums
over
$\ell$ and $r$ in \eq{hibac2}, and then exploit the following formula
$$
\sum_{\ell=1}^n f(\ell) = \sum_{\ell=1}^n \cos\left(\frac{2\pi}{N}\,
\ell \right)
=\frac{1}{2}\left[\frac{\sin\frac{2\pi}{N}\left(n+\frac{1}{2}
\right)}{\sin\frac{\pi}{N}}-1\right]~~~.
$$
After some straightforward algebra, we find
\bea
{\cal H}_{back}^{(2)} &=&
\frac{J_z}{N}\,\frac{1}{\sin\frac{\pi}{N}}
\sum_{r=-\infty}^{\infty}
\sin\frac{2\pi}{N}\left(
r-\gamma+\frac{1}{2}\right)
\left(:a^\dagger_r\,a_r:+:b^\dagger_r\,b_r:\right)\nl
&=&
\frac{J_z}{N}\,\frac{1}{\sin\frac{\pi}{N}}
\sum_{r=-\infty}^{\infty}
\left[\cos\pi\sigma\,\sin\frac{2\pi}{N}\left(
r-\frac{1}{2}\right)\right.
\label{hibac2'}\\
&&+\ \left.\sin\pi\sigma\,\cos\frac{2\pi}{N}\left(
r-\frac{1}{2}\right)\right]
\left(:a^\dagger_r\,a_r:+:b^\dagger_r\,b_r:\right)~~~.
\nonumber
\ena

We now summarize our results by collecting all terms of the effective
hamiltonian ${\cal H}$; for future reference, we
organize them as power series in $1/N$ according to
\beq
{\cal H} = \sum_{k=0}^\infty
\left(\frac{2\pi}{N}\right)^k\,{\cal H}_{(k)}~~~.
\label{series}
\eeq
Using Eqs. (\ref{h0+}),
(\ref{hifor4'}), (\ref{hibac4'}) and (\ref{hibac2'}),
the first few terms of this series are
\beq
{\cal H}_{(0)} = \left(-B+J_z\,\sigma+\sin\frac{\pi\sigma}{2}
+\frac{J_z}{\pi}\,
\sin\pi\sigma\right)\sum_{r=-\infty}^\infty
\left(:a^\dagger_r\,a_r:+:b^\dagger_r\,b_r:\right)~~~,
\label{hof}
\eeq
\bea
{\cal H}_{(1)} &=& \left(\cos\frac{\pi\sigma}{2} +
\frac{J_z}{\pi}\cos\pi\sigma\right) \sum_{r=-\infty}^\infty
\left(r-\frac{1}{2}\right)
\left(:a^\dagger_r\,a_r:+:b^\dagger_r\,b_r:\right)\nl
&&+\ \frac{J_z}{2\pi}\sum_{\ell,r,s=-\infty}^{\infty}
\left(:a^\dagger_{r-\ell}\,a_r:\,
:a^\dagger_{s+\ell}\,a_s:+:b^\dagger_{r+\ell}\,b_r:\,
:b^\dagger_{s-\ell}\,b_s:\right)
\label{h1f} \\
&&+\ \frac{J_z}{\pi}\,(\cos\pi\sigma+1)\sum_{\ell,r,s=-\infty}^\infty
:a^\dagger_{r-\ell}\,a_r:\,:b^\dagger_{s-\ell}\,b_s:~~~,
\nonumber
\ena
\bea
{\cal H}_{(2)} &=&
-\ \frac{1}{2}\left(\sin\frac{\pi\sigma}{2}+
\frac{J_z}{\pi}\sin\pi\sigma\right)\sum_{r=-\infty}^\infty
\left(r-\frac{1}{2}\right)^2
\left(:a^\dagger_r\,a_r:+:b^\dagger_r\,b_r:\right)
\nl
&&+\ \frac{J_z}{24\pi}\sin\pi\sigma
\sum_{r=-\infty}^\infty
\left(:a^\dagger_r\,a_r:+:b^\dagger_r\,b_r:\right)
\label{h2f} \\
&&-\ \frac{J_z}{\pi}\sin\pi\sigma
\sum_{\ell,r,s=-\infty}^\infty
(r+s-\ell-1):a^\dagger_{r-\ell}\,a_r:\,:b^\dagger_{s-\ell}\,b_s: ~~~,
\nonumber
\ena
and
\bea
{\cal H}_{(3)} &=&-\ \frac{1}{6}\left(\cos\frac{\pi\sigma}{2} +
\frac{J_z}{\pi}\cos\pi\sigma\right)\sum_{r=-\infty}^\infty
\left(r-\frac{1}{2}\right)^3
\left(:a^\dagger_r\,a_r:+:b^\dagger_r\,b_r:\right)
\nl
&&+\ \frac{J_z}{24\pi}\cos\pi\sigma
\sum_{r=-\infty}^\infty
\left(r-\frac{1}{2}\right)
\left(:a^\dagger_r\,a_r:+:b^\dagger_r\,b_r:\right)
\label{h3f} \\
&&-\ \frac{J_z}{4\pi}\sum_{\ell,r,s=-\infty}^{\infty}
\ell^2\left(:a^\dagger_{r-\ell}\,a_r:\,
:a^\dagger_{s+\ell}\,a_s:+:b^\dagger_{r+\ell}\,b_r:\,
:b^\dagger_{s-\ell}\,b_s:\right)
\nl
&&-\ \frac{J_z}{2\pi}
\sum_{\ell,r,s=-\infty}^\infty
\left[(r+s-\ell-1)^2
\cos\pi\sigma+\ell^2\right]
:a^\dagger_{r-\ell}\,a_r:\,:b^\dagger_{s-\ell}\,b_s:~~~.
\nonumber
\ena
Notice that ${\cal H}_{(0)}$ and ${\cal H}_{(2)}$ vanish when
$B=0$ (and hence $\sigma=0$). This property actually holds for
all terms ${\cal H}_{(k)}$ with $k$ even. It is also interesting
to observe that the two terms proportional to $J_z$ in ${\cal
H}_{(0)}$
have a different origin. The first comes directly from the ``free''
hamiltonian ${\cal H}_0$ of \eq{h0}, while the second comes
from the backward scattering part of the interaction hamiltonian
and thus is a normal ordering effect
\footnote{ As a useful check, let us observe that the effective
Hamiltonian
${\cal H}$ is invariant under the transformations $B\to -B$, $\sigma
\to -\sigma$, $a_r\to a_{1-r}^\dagger$, $b_r\to b_{1-r}^\dagger$, as
the original hamiltonian (\ref{heisham0}) is invariant under
$B\to -B$ and ${\bf S}_i\to -{\bf S}_i$.}.

We conclude our discussion of the Heisenberg
model with a few more comments.
The effective hamiltonian given in Eqs. (\ref{series})-(\ref{h3f})
is similar but not identical to that of Ref.~\cite{lupe} (where
only the $1/N$-term with a vanishing magnetic field is explicitly
considered). The difference between our results and those
of Ref.~\cite{lupe} is due to a different normal ordering
prescription
for the interaction hamiltonian ${\cal H}_I$. In our derivation, we
have always consistently used the normal ordering
as dictated by the Jordan-Wigner transformation. Consequently,
the forward scattering part given in
\eq{hifor4'}, is automatically
normal ordered in each pair of fermions. The
backward scattering part, instead,
requires a rearrangement that produces
a normal ordered four-fermion piece given in
\eq{hibac4'},
and also a two-body part given in
\eq{hibac2'}. It is precisely the latter
in addition to the forward scattering part
that causes the difference between our
effective hamiltonian and that of Ref.~\cite{lupe}.
We will comment more on this fact in \secn{heisw} when we compare
our results and the exact Bethe Ansatz solution.

\bigskip
\noindent{\bf The Calogero-Sutherland model}

The Calogero-Sutherland model describes the interaction
of $N$ non-relativistic fermions of mass $m$ moving on a circle
of length $L$ with a pairwise potential
proportional to the inverse square of the chord distance between the
two particles \cite{cal,sut}. If we denote by $x_i$ the coordinate of
the $i$-th fermion along the circle and choose units such that $2m=1$,
then the hamiltonian is
\beq
h=-\sum_{j=1}^N \frac{\partial^2}{\partial x_j^2}
+g\,\frac{\pi^2}{L^2}\sum_{j<k}
\frac{1}{\sin^2 (\pi(x_j-x_k)/L) }~~~,
\label{hcs}
\eeq
where $g$ is the coupling constant. In the following, without any
loss
of generality, we will take $N$ to be odd.

This model can be exactly solved by Bethe Ansatz and all its
fundamental
properties can be obtained from this
solution \cite{sut}. In particular, one finds
that the low-energy excitations above the ground state are
gapless, and thus the long distance properties of the system are
described by a conformal field
theory \cite{kaya}. In what follows, we will derive
explicitly the effective theory of the Calogero-Sutherland model
and show that it is identical in structure
to that of the Heisenberg model.

The single particle wave functions of $h$ are plane waves
$$
\phi_n(x) = \frac{1}{\sqrt{L}}\ \exp\left({\ri} \frac{2\pi}{L}
nx\right)~~~,
$$
with $n \in {\bf Z}$ to satisfy periodic boundary conditions.
Introducing a set of fermionic oscillators with standard
anticommutation
relations (cf. \eq{psipsi}), we
can define a second quantized non-relativistic fermion field as
$$
\Psi(x) = \sum_{n=-\infty}^\infty \psi_n \,\phi_n(x)~~~,
$$
and then compute the second quantized hamiltonian
corresponding to $h$, which is given by
$$
H = H_0 + H_I~~~,
$$
where the kinetic part is
\bea
H_0 &=&\int^L_0 dx~\Psi^{\dagger}(x)\left(-
\frac{\partial^2}{\partial x^2}\right) \Psi(x) \nl
&=& \left(\frac{2\pi}{L}\right)^2\sum_{n=-\infty}^\infty
n^2\,\psi_n^{\dagger}\,\psi_n~~~,
\label{h0cs}
\ena
while the interaction part is \cite{clz}
\bea
H_I &=&\frac{1}{2} \int^L_0 dx  \int^L_0 dy ~
\Psi^{\dagger}(x)\,\Psi^{\dagger}(y)\left[
g\,\frac{\pi^2}{L^2}
\frac{1}{\sin^2 (\pi(x -y)/L) }\right]
\Psi(x)\,\Psi(y) \nl
&=&- \ g \,\frac{\pi^2}{L^2} \sum_{l,n,m=-\infty}^\infty |l|
{}~\psi_{m+l}^{\dagger}\,\psi_{n-l}^{\dagger}\,
\psi_n \,\psi_m ~~~.
\label{hics}
\ena
The hamiltonian $H_0$ simply describes a free fermionic system whose
ground state $|\Omega\rangle$ is given by \eq{ground} with
\beq
n_F = \frac{N-1}{2} ~~~.
\label{nfcs}
\eeq

To study the small fluctuations around the Fermi points $\pm n_F$, we
define two independent sets of oscillators, $a_r$ and $b_r$,
according to \eq{arbr}.
In terms of these degrees of freedom, the kinetic part of the
hamiltonian
reads
\beq
{\cal H}_0=
\left(\frac{2\pi}{L}\right)^2\
\sum_{r=-\Lambda_0}^{\Lambda_0}\
\left(n_F+r\right)^2 \,
\left(:a^\dagger_r\,a_r:+:b^\dagger_r\,b_r:\right)~~~,
\label{kinab}
\eeq
where the bandwidth cut-off is such that $\Lambda_0<<N$ and
$\Lambda_0=o(N)$ in the thermodynamic limit $L,N\to\infty$, as in
the Heisenberg model.

Let us now consider the interaction hamiltonian $H_I$.
Our purpose is to write the corresponding effective operator
${\cal H}_I$ in terms of the usual
bilinear fermionic forms; once this is done,
it is simple to interpret the results within the algebraic
context of the extended conformal theories.
To do so, however, we need to
reorder the oscillators of $H_I$, for example by moving
$\psi_m$ close to $\psi^\dagger_{m+\l}$ in \eq{hics}.
In this way, we obtain a four-fermion term that is completely
similar in structure to
the interaction hamiltonian of the Heisenberg model (cf. \eq{hi}),
and
thus can be treated accordingly. The
price we pay to achieve this is the appearance of a divergent
two-fermion
term. Thus, before we can proceed, it is necessary to
introduce a regularization prescription to avoid this divergence and
give a meaning to
our formulas. Inspired by the procedure followed in our discussion of
the Heisenberg model, we perform a ``periodic regularization'',
namely
we divide the momentum space into ficticious Brillouin
zones with amplitude $2M$, {\it i.e.} we identify $l$ and $l'$
if $l'=l+2M\,k$ for any integer $k$.
The number $M$, serving as a regulator, is taken to be arbitrary
with the only constraint $M >> n_F$ such 
that the physical region of interest is inside the first zone.
Thus, we can perform all calculations involving small oscillations
around the Fermi points
$\pm n_F$ in the first Brillouin zone, like in the Heisenberg model,
and, at the end, let $M\to\infty$ to recover the original continuum
theory. We will now show that this procedure
is consistent and leads to finite and meaningful results.

According to this regularization prescription,
the hamiltonian $H_I$ of \eq{hics} becomes
$$
H_I(M) = H_I'(M)+H_I''(M)
$$
where
\bea
H_I'(M) &=& -\ g\,\frac{\pi^2}{L^2}
\sum_{n,n',m,m'=-M+1}^M
\big|n-n'\big|_M
\,:\psi_{n'}^\dagger\,\psi_n:~:\psi_{m'}^\dagger\,\psi_m:
\nl
&&\times\ \delta(n'-n+m'-m)~~~,
\label{him'}
\ena
and
\beq
H_I''(M)= g\,\frac{\pi^2}{L^2}
\left(\sum_{l=-M+1}^{M}
\big|l\big|_M\right) \sum_{n=-M+1}^M :\psi_{n}^{\dagger}\,\psi_n:~~~.
\label{him"}
\eeq
In these equations, all indices have been restricted to the
first Brillouin zone and consequently, the absolute value has been
replaced
by its periodic extension modulo $2M$, which we
have denoted by the symbol
$\,\big|~\big|_M$. To be specific, we have
\beq
\big|\kappa\big|_M ~=~ |\kappa|~~~,~~~
\big|M+\kappa\big|_M ~=~ \big|-M+\kappa\big|_M=M-|\kappa|
\label{mod}
\eeq
for $|\kappa|\leq M$.

The four-fermion part $H_I'(M)$ is identical in form to the first
line of \eq{hi}, and hence the corresponding
effective hamiltonian ${\cal H}_I'(M)$ contains
only a forward and a backward scattering part
\footnote{Of course, the Umklapp terms are
neither present in the original continuum model, nor
produced by our periodic regularization.},
namely
$$
{\cal H}_I'(M)={\cal H}_{forw}(M)+ {\cal H}_{back}(M)
$$
The forward scattering terms arise when the indices in \eq{him'}
satisfy the conditions (\ref{confforw}) and (\ref{ranges}), so that
\bea
{\cal H}_{forw}(M) &=&  - \ g\,\frac{\pi^2}{L^2}~
{ \sum_\ell}' \sum_{r,s=-\Lambda_0}^{\Lambda_0} \big|\ell\big|_M
\left[:a^\dagger_{r-\ell}\,a_r:\,:a^\dagger_{s+\ell}\,a_s:\right.
\label{hifcs}\\
&&+\ \left. :b^\dagger_{r+\ell}\,b_r:\,
:b^\dagger_{s-\ell}\,b_s:
+:a^\dagger_{r-\ell}\,a_r:\,:b^\dagger_{s-\ell}\,b_s:
+:b^\dagger_{r+\ell}\,b_r:\,:a^\dagger_{s+\ell}\,a_s:\right]~~~.
\nonumber
\ena
The backward scattering terms, instead, appear when the indices in
\eq{him'} satisfy the conditions (\ref{confback}) and
(\ref{ranges}), so that
\bea
{\cal H}_{back}(M)&=& -\ g\,\frac{\pi^2}{L^2}~
{ \sum_\ell}' \sum_{r,s=-\Lambda_0}^{\Lambda_0}
 \left[
\big|2n_F+\ell \big|_M\,
b^\dagger_{-r+\ell}\,a_r\,a^\dagger_{-s+\ell}\,b_s \right. \nl
&&+\ \left.\big|-2n_F+\ell \big|_M\,
a^\dagger_{-r-\ell}\,b_r\,b^\dagger_{-s-\ell}\,a_s \right]~~~.
\label{hibcs}
\ena

Finally, let us consider the two-fermion term $H_I''$ of \eq{him"}.
Using
the effective degrees of freedom, it becomes
\beq
{\cal H}_I''(M) =
g\,\frac{\pi^2}{L^2}\,M^2\sum_{r=-\Lambda_0}^{\Lambda_0}
\left(:a^{\dagger}_r\, a_r :+:b^{\dagger}_r\, b_r:\right)
{}~~~.
\label{hidue}
\eeq
Of course, in the limit $M\to\infty$, ${\cal H}_I''(M)$ is divergent.
However, when one considers the full hamiltonian, this divergence
disappears.

To see this, and also to establish a correspondence
with the extended conformal theories, we must remove
the bandwidth cut-off.
Thus, following the same strategy (and using the same conventions)
of the Heisenberg model, we send $N \to\infty$ and
$\Lambda_0\to\infty$ in all previous formulas.
In this limit, using \eq{nfcs}, we easily
see that \eq{kinab} becomes
\beq
{\cal H}_0=\left(2\pi\rho_0\right)^2 \sum_{r=-\infty}^\infty
\left[
\frac{1}{4} + \frac{1}{N} \left(r-\frac{1}{2}\right) +
\frac{1}{N^2}\left(r-\frac{1}{2}\right)^2 \right]
\left(:a^{\dagger}_r \,a_r:+:b^{\dagger}_r \,b_r:\right)~~~,
\label{kindue}
\eeq
where $\rho_0 =N/L$ is the density, which is kept fixed in the
thermodynamic limit. As in the Heisenberg case, ${\cal H}_0$ is
meaningful
only when acting on low-energy states with particle and hole momenta
bounded by $\Lambda_0<<N$.

Analogously, the forward scattering terms, \eq{hifcs}, simply turn
into
\bea
{\cal H}_{forw}(M) &=& -\ \frac{g}{4}\left(2\pi \rho_0\right)^2
\frac{1}{N^2}
\sum_{\ell,r,s=-\infty}^\infty|\ell|
\left[:a^\dagger_{r-\ell}\,a_r:\,:a^\dagger_{s+\ell}\,a_s:\right.
\nl
&&+\ \left. :b^\dagger_{r+\ell}\,b_r:\,:b^\dagger_{s-\ell}\,b_s:
+2\,:a^\dagger_{r-\ell}\,a_r:\,:b^\dagger_{s-\ell}\,b_s: \right]~~~.
\label{hif1cs}
\ena
Notice that ${\cal H}_{forw}$ is independent of the regularization
parameter $M$, and is similar in structure to the forward scattering
hamiltonian of the Heisenberg model (cf. \eq{hifor4'}).

The backward scattering terms \eq{hibcs} require, instead, more care
since a reordering of the $a$ and $b$ oscillators is necessary. 
Before doing this rearrangement, we remove the
bandwidth cut-off and use the definition (\ref{mod}) of the
periodic modulus to rewrite ${\cal H}_{back}$ as follows
\bea
{\cal H}_{back}(M) &=& -\ \frac{g}{4} \left(2\pi \rho_0\right)^2
\frac{1}{N^2}
\sum_{\ell,r,s=-\infty}^\infty\left(M - |\ell|\right)
\left[
b^\dagger_{-r+\ell+\gamma}\,a_r\,a^\dagger_{-s+\ell+\gamma}\,b_s
\right.\nl
&&+\
\left.a^\dagger_{-r-\ell+\gamma}\,b_r\,b^\dagger_{-s-\ell+\gamma}\,a_s
\right]~~~,
\label{hibga}
\ena
where
\beq
\gamma = M-N+1 = M - 2n_F~~~.
\label{gamma'}
\eeq
The two terms in the square bracket of \eq{hibga} describe two
backward scattering processes that exchange the same momentum. In
fact,
$(2n_F+\ell+\gamma)$, which is the momentum exchanged by
the first term, and $(-2n_F+\ell-\gamma)$, which is the momentum
exchanged by
the second term, differ by $2M$, {\it i.e.} by a period.
At this point we can proceed in close analogy to the steps that
broughts us
from \eq{hibac1} to Eqs. (\ref{hibac4'}) and (\ref{hibac2'}), namely
we normal order ${\cal H}_{back}(M)$ without breaking the
square bracket. After some straightforward algebra, we find
$$
{\cal H}_{back}(M)={\cal H}_{back}^{(4)}(M)+{\cal H}_{back}^{(2)}(M)
$$
where
\bea
{\cal H}^{(4)}_{back}(M) &=& \frac{g}{2}\left(2\pi
\rho_0\right)^2
\frac{1}{N^2}\nl
&&\times\ \sum_{r,s=-\infty}^\infty
\left[
\sum_{\ell=-\infty}^{r+s-M+N-1}\left( 2M-N-r-s +\ell+1\right)
:a^\dagger_{r-\ell}\,a_r:\,:b^\dagger_{s-\ell}\,b_s:
\right. \nl
&&+\ \left.
\sum_{\ell=r+s-M+N}^\infty\left( N+r+s -\ell-1\right)
:a^\dagger_{r-\ell}\,a_r:\,:b^\dagger_{s-\ell}\,b_s:\right]~~~,
\label{hquatg}
\ena
and
\bea
{\cal H}^{(2)}_{back}(M) &=& \frac{g}{4}\ \left(2\pi
\rho_0\right)^2 \frac{1}{N^2}\Bigg\{
-M^2\sum_{r=-\infty}^\infty  \left( : a^{\dagger}_r \,a_r:+
:b^{\dagger}_r\, b_r:\right)\nl
&&+\
\sum_{r=-\infty}^\infty
\left[ N^2 + N(2r-1) + r(r-1) \right]
 \left( : a^{\dagger}_r \,a_r:+:b^{\dagger}_r\, b_r:\right)
\label{hdueg}\\
&&+\ 2 \sum_{r=M-N+1}^\infty
\left( r-M+N \right) \left( r-M+N-1)\right)
\left( :a^{\dagger}_r\, a_r:+:b^{\dagger}_r \,b_r:\right)\Bigg\}~~~.
\nonumber
\ena
Since these terms depend explicitly
on the regularization parameter $M$, we must check
that they combine to give a finite result when $M\to\infty$.
To this aim, let us consider first the four-fermion operators given
by
Eqs. (\ref{hif1cs}) and (\ref{hquatg}). We have already
remarked that ${\cal H}_{forw}(M)$ is actually independent of $M$; on
the contrary ${\cal H}^{(4)}_{back}(M)$ does depend on $M$ but,
in the limit $M\to\infty$, it reduces to
\beq
{\cal H}^{(4)}_{back} = \frac{g}{2}\left(2\pi \rho_0\right)^2
\frac{1}{N^2}\sum_{\ell,r,s=-\infty}^\infty
(N+r+s-\ell-1)
\,:a^\dagger_{r-\ell}\,a_r:\,
:b^\dagger_{s-\ell}\,b_s:~~~.
\label{ht}
\eeq
Indeed, the first term in the r.h.s. of
\eq{hquatg} vanishes when $M\to\infty$, because, given any two
states $|v\rangle$ and $|w\rangle$ of the fermionic Fock space,
there exists always a positive number $k$ such that
$$
\langle
v|:a^\dagger_{r-\ell}\,a_r:\,:b^\dagger_{s-\ell}\,b_s:|w\rangle
=0
$$
for any $\ell<-k$.
For the same reason, the last term in the r.h.s. of \eq{hdueg}
vanishes when $M\to\infty$. Moreover, the first term
exactly cancels with  ${\cal H}_I''(M)$ of \eq{hidue}.
Therefore, if we combine ${\cal H}^{(2)}_{back}(M)$
with ${\cal H}_I''(M)$ we get a finite result. Indeed,
\bea
{\cal H}^{(2)} &\equiv& \lim_{M \to \infty} \left(
{\cal H}_I''(M) + {\cal H}^{(2)}_{back}(M)\right)
\label{hdos}\\
&=&
\frac{g}{4}\left(2\pi \rho_0\right)^2
\frac{1}{N^2}
\sum_{r=-\infty}^\infty \left[N^2 + N(2r-1) + r(r-1) \right]
\left( : a^{\dagger}_r \,a_r:+
:b^{\dagger}_r\, b_r:\right)~~~.
\nonumber
\ena

We now summarize our results by writing the complete
effective hamiltonian of the Calogero-Sutherland model. This is
given by
\beq
{\cal H} = \left(2\pi \rho_0\right)^2
\sum_{k=0}^2 \frac{1}{N^k}\,{\cal H}_{(k)}
\label{seriesc}
\eeq
where
\beq
{\cal H}_{(0)} =\frac{1}{4}\,(1+g)\sum_{r=-\infty}^\infty
\left( : a^{\dagger}_r \,a_r:+
:b^{\dagger}_r\, b_r:\right)~~~,
\label{hc0}
\eeq
\bea
{\cal H}_{(1)} &=&\left(1+\frac{g}{2}\right)
\sum_{r=-\infty}^\infty  \left(r-\frac{1}{2}\right)
\left( : a^{\dagger}_r \,a_r:+
:b^{\dagger}_r\, b_r:\right)\nl
&&+\ \frac{g}{2}\sum_{\ell,r,s=-\infty}^\infty
:a^\dagger_{r-\ell}\,a_r:\,
:b^\dagger_{s-\ell}\,b_s:~~~,
\label{hc1}
\ena
and
\bea
{\cal H}_{(2)} &=&
\sum_{r=-\infty}^\infty  \left[\left(r-\frac{1}{2}\right)^2
+\frac{g}{4}\left(r^2-r\right)\right]
\left( : a^{\dagger}_r \,a_r:+
:b^{\dagger}_r\, b_r:\right)\nl
&&-\ \frac{g}{4}
\sum_{\ell,r,s=-\infty}^\infty|\ell|
\left[:a^\dagger_{r-\ell}\,a_r:\,:a^\dagger_{s+\ell}\,a_s:
+:b^\dagger_{r+\ell}\,b_r:\,:b^\dagger_{s-\ell}\,b_s: \right.\nl
&&~~~+\left.2\,:a^\dagger_{r-\ell}\,a_r:\,:b^\dagger_{s-\ell}\,b_s:
\right]
\label{hc2}\\
&&+\ \frac{g}{2}
\sum_{\ell,r,s=-\infty}^\infty(r+s-\ell-1)
:a^\dagger_{r-\ell}\,a_r:\,:b^\dagger_{s-\ell}\,b_s: ~~~.
\nonumber
\ena
Notice that there are no contributions to
${\cal H}$ of order $1/N^3$ or higher.

We conclude this section with a few remarks. As in the
Heisenberg case, also here
the backward scattering part of the interaction hamiltonian
plays a crucial role. In fact, in spite
of the overall factor $1/L^2$ in \eq{hics}, it contributes
both to the zeroth- and first-order effective hamiltonians
${\cal H}_{(0)}$ and ${\cal H}_{(1)}$, due to normal ordering
effects. On the contrary, the forward scattering part
contributes only to ${\cal H}_{(2)}$. In \secn{csw} we will
interpret these
results in the context of the $\winf$algebra.

Let us now comment on the general validity of our approach.
Both the Heisenberg and Calogero-Sutherland models
lead to effective hamiltonians with the same operator structure,
which only differ in the functional form of the dispersion and
scattering terms. In fact, our method would be equally valid for any
other fermionic model of the same form, namely for a hamiltonian
with a gapless bilinear kinetic term and
a four-fermion interaction term. This class of models is known as
the class of generalized Luttinger systems
\cite{hald1,fermi,kaya,clz}.
Note that in our approach there are no special restrictions on the
specific form of the dispersion and scattering functions.
In particular, no reference to integrability is ever made.

\vskip 1.5cm
\section{The $\winf$algebraic approach}
\label{w}
\bigskip

In the previous section we have seen that the low-energy
effective hamiltonian both for the Heisenberg
and the Calogero-Sutherland models can be written entirely in terms
of
fermionic bilinear operators of the generic form
$$
{\cal O}_\ell = \sum_{r=-\infty}^\infty
f_\ell(r)\,:a^{\dagger}_{r-\ell}\, a_r : ~~~,
$$
or
$$
{\overline {\cal O}}_\ell = \sum_{r=-\infty}^\infty
f_\ell(r)\,:b^{\dagger}_{r-\ell}\, b_r : ~~~.
$$
In this section we shall exhibit a very natural basis of
operators ${\cal O}_\ell$ and ${\overline{\cal O}}_\ell$
which satisfy the infinite dimensional algebra known
as $\winf$\cite{shen,kac1}.
In practice, this simply amounts to recognize a special basis
of polynomials for the functions $f_\ell(r)$.
However, this is not merely a change of basis. Indeed, by displaying
the $\winf$structure of the theory one can take advantage of the fact
that this algebra can be realized also by bosonic
operators. This means in particular, that once
the algebraic content of the
fermionic theory has been established, other realizations
of the {\it same} algebra can be constructed in the bosonic language,
and these can be chosen to diagonalize the total hamiltonian.
For this
reason, we shall call this procedure {\it algebraic bosonization}.
Before proceeding to rewrite the results of the previous
section in terms of this new basis, we shall briefly review the
essentials
of a generic theory based on the $\winf$algebra.

\bigskip
\noindent{\bf The $\winf$algebra}

The low-energy dynamics of simple $(1+1)$-dimensional fermionic
systems
can be described by the (small) fluctuations of the zero-dimensional
Fermi ``surface'' (which actually consists of an
even number of Fermi points) around the one-dimensional Fermi sea.
The effective degrees of freedom
describing these fluctuations are the bosonized variables of the
underlying fermionic theory (for formulations of
bosonization that are close to our ideas see, {\it e.g.},
Refs.~\cite{hald1,froh,boson}). A systematic way
of studying these fluctuations is to recognize their characteristic
dynamical symmetry and organize them into irreducible representations
of it. In Ref.~\cite{clz} this dynamical symmetry has been identified
as the infinite dimensional $\winf$algebra \cite{shen}.
More precisely, for those systems in which parity is unbroken,
for example in the Heisenberg and the Calogero-Sutherland models, the
dynamical symmetry is given by the algebra
$\winf \times \wb_{1+\infty}$, where
each factor is associated to each
Fermi point. Specifically, $\winf$is
the chiral symmetry algebra of the right Fermi point whereas
$\wb_{1+\infty}$ is the antichiral symmetry algebra of the left
Fermi point.
In most of the discussion that follows in this Section, it will
be enough to consider, without loss of generality, only
one component, say the chiral one; however
we will point out explicitly all cases in which
the combination of the chiral and
antichiral sectors is relevant.

In simple terms, the (chiral) $\winf$symmetry is an extension of the
usual
(chiral) conformal symmetry of $(1+1)$-dimensional relativistic
systems with
massless excitations.
Here, the relativistic ``massless'' excitations are the
small fluctuations of the fermions close to the Fermi points in
the momentum space of the non-relativistic system.
One can imagine that the extended conformal symmetry
has its origin precisely
in the corrections to the approximate linear dispersion
law around the Fermi points that the system exhibits.
Moreover, as we shall see by explicit construction, the inclusion
of the extra generators of the enhanced symmetry
allows us to produce a systematic $1/N$ expansion in a very natural
way.

The $\winf$algebra is generated by an infinite set of (chiral)
currents
$V^i_n$, which are characterized by a (momemtum) mode index
$n \in {\bf Z}$ and an integer conformal spin $h =i+1 \ge 1$.
Roughly speaking, the geometrical meaning of the index $i$ is
associated
to the type of ``multipole'' deformation the current $V^i_n$
can induce on physical states.
These currents satisfy the algebra \cite{shen},
\beq
{[\ V^i_n, V^j_m\ ]} = (jn-im) V^{i+j-1}_{n+m}
+q(i,j,n,m)V^{i+j-3}_{n+m}
+\cdots +\delta^{ij}\delta_{n+m,0}\ c\ d(i,n) \ ,
\label{walg}
\eeq
where the structure constants $q(i,j,n,m)$ and $d(i,n)$ are
polynomial
in their arguments, $c$ is the central charge, and the dots denote a
finite number of terms involving the operators $V^{i+j-2k}_{n+m}\ $
(the complete expression of \eq{walg} is a bit cumbersome
and is given in the Appendix).

In particular,
the operators $V^0_n$ satisfy the Abelian current algebra (Kac-Moody
algebra)
$\u1$, while the operators $V^1_n$ close the Virasoro algebra
\cite{bpz},
that is
\barr
{[\ V^0_n,V^0_m\ ]}\    & = & c\ n\ \delta_{n+m,0} ~~~,
\label{walg0} \\
{[\ V^1_n, V^0_m\ }]\ & = & -m\ V^0_{n+m} ~~~,\label{walg01}\\
{[\ V^1_n, V^1_m\ ]}\ & = & (n-m)V^1_{n+m} +{c\over 12}n(n^2-1)
\delta_{n+m,0}~~~.
\label{walg1}
\earr
The eigenvalues of $V^0_0$ and $V^1_0$ are identified, respectively,
as the charge and conformal dimension
of a chiral excitation.

Other algebraic relations contained in \eq{walg},
which will be useful in the next sections,
are
\barr
\left[\ V^2_n, V^0_m\ \right] &=& -2m\ V^1_{n+m}~~~,\nonumber\\
\left[\ V^2_n, V^1_m\ \right] &=& (n-2m)\ V^2_{n+m} -
     {1\over 6}\left(m^3-m\right) V^0_{n+m}~~~,\nonumber\\
\left[\ V^2_n, V^2_m\ \right] &=& (2n-2m)\ V^3_{n+m}
     +{n-m\over 15}\left( 2n^2 +2m^2 -nm-8 \right) V^1_{n+m}
\nonumber\\
     &&\quad +\ c\ {n(n^2-1)(n^2-4)\over 180}\ \delta_{n+m,0}~~~.
\label{com}
\earr
It is interesting to note that the operators $V_0^i$ commute with
each other
for any $i$, and thus are the generators of the Cartan subalgebra of
$\winf$.

In the classical limit, all terms but the first in the r.h.s. of
\eq{walg} vanish; the resulting algebra is the classical algebra
$w_\infty$ of area-preserving diffeomorphisms, which can be
understood
as originating from all classical deformations of the density
which conserve the number of particles \cite{clz}.

A chiral $\winf$theory is defined by a Hilbert space constructed
out of a set of irreducible, unitary, highest-weight \reps of the
$\winf$algebra,
which are closed under the fusion rules for making composite states.
If the parity symmetry is unbroken, the complete Hilbert space
is obtained by combining chiral and antichiral representations
of $\winf$. This is a simple extension of the well-known construction
of conformal field theories. For this reason, and also because
$\winf$contains
the Virasoro algebra as a subalgebra, a $\winf$theory is
called an extended conformal field theory.

All such theories can be completely classified
thanks to the crucial work of Kac and Radul
\cite{kac1},
in which all irreducible, unitary, quasi-finite highest-weight
\reps of (chiral) $\winf$have been constructed.
Such representations exist only if the central charge is a positive
integer, {\it i.e.}  $c\in {\bf Z}_+$ \footnote{
For the special case of the Luttinger systems, like the Heisenberg
and the Calogero-Sutherland models described in this paper,
we have $c=1$ \cite{kaya,clz}.}.
They
are characterized by an $c$-dimensional weight vector
$\vec{Q}$ with real elements, and are built on top of
a highest weight state $|\,\vec{Q}\,\rangle$, which satisfies
\beq
V^i_n\,|\,\vec{Q}\,\rangle=0
\label{whst}
\eeq
for any $n>0$ and $i \ge 0$,
and
\beq
V^i_0\,|\,\vec{Q}\,\rangle = \sum_{\alpha=1}^c \ m^i (Q_\alpha)\
|\,\vec{Q}\,\rangle ~~~,
\label{weig}
\eeq
where $m^i(Q)$ are $i$-th order polynomials.
In particular, for $i=0,1,2$,
\bea
m^0(Q) &=& Q~~~,
\nl
m^1(Q) &=& \frac{1}{2}\,Q^2~~~,
\label{wcs} \\
m^2(Q) &=& \frac{1}{3}\,Q^3~~~.
\nonumber
\ena
The eigenvalue of $V_0^0$ is the charge
(defined by the $\u1$ symmetry of local particle number
conservation) of $|\,\vec{Q}\,\rangle$,
while that of $V_0^1$ gives its conformal weight.

The complete highest weight representation (the so-called Verma
module) is obtained by constructing
all the descendant states of $|\,\vec{Q}\,\rangle$. These
correspond to neutrally charged (particle-hole) excitations
above $|\,\vec{Q}\,\rangle$, and
are defined as follows
\beq
|\,\vec{Q}\, ,\,\{ k_i \}\,\rangle=
V^0_{-k_1}\ V^0_{-k_2} \cdots V^0_{-k_s } \,|\,\vec{Q}\,\rangle
{}~~~,\quad k_1 \ge k_2\ge \cdots \ge k_s > 0~~~.
\label{neut}
\eeq
The quantity $k=\sum\limits_{i=1}^{s} k_i$
represents the total momentum of
the excitation measured with respect to $|\,\vec{Q}\,\rangle$, and
it is also known as the {\it level} of the
descendant state $|\,\vec{Q}\, ,\,\{ k_i \}\,\rangle$.

We conclude this brief survey by pointing out that the effective
theory
of a Luttinger system \cite{hald1,fermi}
is a $W_{1+\infty} \times \wb_{1+\infty}$ conformal theory with
$c={\bar c}=1$ \cite{clz}.
In this case the highest weight vector $\vec{Q}$ is one-dimensional
and the corresponding highest weight states are
denoted by $|\,Q\,\rangle$ in the right
and $|\,{\overline Q}\,\rangle$ in the left sectors,
with $Q$ and ${\overline Q}$ being their respective charges.
Later on, we will show that this theory can be realized both by
a free fermion and by a free compactified boson.

\bigskip
\noindent{\bf The Weyl fermion realization of $\winf$}

In \secn{linear} we have seen that
the effective hamiltonian of the $(1+1)$-dimensional
systems we considered, is constructed entirely in terms
of the chiral and antichiral relativistic Weyl fermions
(\ref{f+}) and (\ref{f-}). We now focus on one of them,
say the chiral one, which is known to describe
a $c=1$ conformal theory.
The Fermi sea for this theory is given by a highest-weight state
$|\,\Omega\,\rangle$ satisfying
\beq
\v^i_n\, |\,\Omega\,\rangle = 0
\label{whwc}
\eeq
for all $n \ge 0$ and $i \ge 0\,$.

The {\it neutrally charged} particle-hole excitations above the
ground state
are described by the descendant states of $|\,\Omega\,\rangle$ (see
\eq{neut}). However, it is
also natural to consider {\it charged excitations}, which
manifest themselves as an excess or a defect of charge around the
Fermi
point.
When considering the complete chiral-antichiral theory, one realizes
that there are two physically inequivalent ways of producing these
excitations: by addition or subtraction of extra particles
with momentum close to the Fermi points,
and by the coupling of the system to an external probe producing an
overall shift in momentum which conserves the total particle number
\cite{kaya,clz}.
The latter are the analogs of Laughlin's quasi-particle
excitations in the quantum Hall
effect \cite{laugh}, as seen from the edge of a sample.
The analogy is evident because Laughlin's quantum incompressible
fluids are configuration space analogs of a Fermi sea.
In the algebraic formalism,
these excitations appear as further highest-weight states,
which have non-vanishing eigenvalues for all the $\v^i_0$ and
$\vb^i_0$.

Therefore, the Hilbert space of a Weyl
fermion consists of an infinity of
$c=1$ $\winf$representations, Eqs. (\ref{whst}) and (\ref{weig}),
which are characterized by an {\it integer} weight representing the
charge of
the highest weight state
\footnote{
The more general representations
with $c=m$ a positive integer can be obtained by considering $m$
independent Weyl fermions (see \cite{ctz5} for details).}.
Obviously, the Hilbert space of an antichiral
fermion is isomorphic to that of a chiral one, and
defines a ${\overline c}=1$ ${\overline \winf}$conformal field
theory.
Therefore, we are led to characterize the thermodynamic limit of
a fermionic Fermi system in which parity is unbroken,
as a $W_{1+\infty} \times {\overline \winf}$theory.

In physical applications, operators and fields are
naturally defined on a spatially compact space, like the circle
of radius $R$ of \eq{dirac}. The corresponding Minkowskian theory is
then defined
on the cylinder formed by the spatial circle times the real line
that represents the time coordinate.
On the other hand, in the mathematical literature operators and
fields are conventionally defined on the complex plane.
Therefore, it is convenient to map the physical operators from
the cylinder to the plane, where one may use
the mathematical results.
There is a well-known conformal mapping between the cylinder
$(u=\tau -iR\theta)$, and the conformal plane $(z)$,
namely
\beq
z=\exp\left({u\over R}\right)=
\exp\left({\tau\over R} -i\theta \right) ~~~,
\label{cft}
\eeq
where $\tau$ denotes the euclidean time.
Under this map, the Weyl fermion (\ref{f+}), which is
a primary field of weight $h=1/2$, takes the form (at $\tau=0$)
\barr
F(z) &=& \left({du\over dz}\right)^{1/2}  F_{+}(\theta) =
\sum_{r=-\infty}^{\infty} {\rm e}^{i\, r\,\theta}\ a_r ~~~, \nl
F^\dagger(z) &=& \left({du\over dz}\right)^{1/2}
F^\dagger_{+}(\theta)
=\sum_{r=-\infty}^{\infty} {\rm e}^{-i\,(r-1)\,\theta}\
a^\dagger_r ~~~.
\label{weypl}
\earr
Notice the well-known fact that due to the map (\ref{cft}), the
definition of $F^\dagger(z)$ differs from the naive expression.
The expression for the antichiral fields can be simply obtained from
\eq{weypl} by replacing $a_r$ with $b_r$, and $\theta$ with
$-\theta$,
{\it i.e.} $z$ with ${\overline z}$.

The representation of the $\winf$generators as operators acting
on the Hilbert space of a Weyl fermion is
discussed in detail in the Appendix. Here we simply recall that it is
obtained by sandwiching specific polynomials
$g^i_n=(-1)^{i+1}f^i_n$ in $D\equiv z\partial$
(with $f^i_n$ given in the Appendix)
between the field operators $F(z)$ and $F^\dagger(z)$, according to
\beq
\v^i_n
= \oint {dz\over 2\pi {\rm i}}\ : F^\dagger(z)\,
z^n\, g^i_n(D) \, F(z):~~~.
\label{defwf}
\eeq
Here the integration is carried clockwise over the unit circle, and
the normal ordering $\ :~~:\ $ is defined canonically
with respect to the ground
state $|\,\Omega\,\rangle$ as in \secn{linear}.
Since the anticommutator of $F$ and $F^\dagger$
is a delta function in Fock space, the operators $\v^i_n$ defined
above clearly represent the $\winf$algebra as long as the
functions $f^i_n$ do so.
One can also verify that the zero modes $\v_0^i$ defined
in this way have the eigenvalues (\ref{wcs}) when
acting on fermion states of charge $Q$.
Furthermore, \eq{defwf} shows how these
operators are written in the canonical form
$(F^\dagger\,F)$ of quantum field theory.

Using the explicit formulae for $f^i_n$ given in the Appendix,
we obtain the Fock space expressions of the generators for
the chiral sector (of course, analogous formulae hold for
the antichiral sector replacing $a_r$ with $b_r$).
For the first few values of the conformal
spin, these are
\barr
\v^0_n &=& \sum_{r=-\infty}^{\infty}: a^\dagger_{r-n}\, a_r :
{}~~~,\nl
\v^1_n &=& \sum_{r=-\infty}^{\infty} \left(\,r-{n+1\over 2}\,\right)
:  a^\dagger_{r-n}\, a_r :~~~,\label{fockw}\\
\v^2_n &=& \sum_{r=-\infty}^{\infty} \left(\,r^2 -(n+1)\ r +
{{(n+1)(n+2)}\over 6}\,
\right) :  a^\dagger_{r-n}\, a_r :~~~,\nl
\v^3_n &=& \sum_{r=-\infty}^{\infty} \left(\,r^3
-{\frac{3}{2}}(n+1)\ r^2 +
{\frac{6n^2+15n+11}{10}}\ r
\right.\nl
&&\qquad\qquad\left. -\ {\frac{(n+1)(n+2)(n+3)}{20}}\,
\right) :  a^\dagger_{r-n}\, a_r :~~~.
\nonumber
\earr
The corresponding currents on the complex plane, {\it i.e.}
$$
V^i (z) \equiv \sum_{n=-\infty}^\infty \v^i_n\ z^{-n-i-1}~~~,
$$
can then be written as follows
\barr
V^0(z) &=& :F^\dagger\,F: ~~~,\nl
V^1(z) &=& {1\over 2}\, :\, \partial \left(F^\dagger\,F\right):
-:F^\dagger\,\partial F: ~~~,\label{wcurp}\\
V^2(z) &=& {1\over 6}\, :\,\partial^2 \left(F^\dagger\,F\right):
- :\partial F^\dagger\,\partial F:~~~, \nl
V^3(z) &=& {1\over 24}\, :\, \partial^3 \left(F^\dagger\,F\right):
 - \,\frac{1}{2}\left(\,:\partial F^\dagger\,\partial^2 F: -
:\partial^2 F^\dagger\,\partial F:\, \right)~~~.
\nonumber
\earr
These expressions will be extensively used in the next sections.

\bigskip
\noindent{\bf The boson realization of $\winf$}

There is yet another realization of the $c=1$ $\winf$algebra
which will be useful for our purposes.
It arises through the abelian bosonization of the fermionic
fields $F(z)$ and $F^\dagger(z)$ introduced in
the previous paragraph.
To see this let us consider a chiral boson field
$\varphi (z)$ defined on the complex plane such that
$$
F(z)\,\equiv \ :\,\exp ( {\rm i}\varphi(z)):
{}~~~,~~~
F^\dagger(z)\, \equiv\ :\,\exp (- {\rm i}\varphi(z)):~~~.
$$
Then one can show that the operator $\partial \varphi(z)$
is a chiral current of conformal weight $(1,0)$,
which is identified
with the lowest spin $\winf$generator $V^0(z)$.
The Fourier expansion of this operator is given by
\beq
\partial \varphi(z) = \sum_{n=-\infty}^\infty \alpha_n\ z^{-n-1}~~~,
\label{fouri}
\eeq
with $\alpha_n \equiv \v^0_n$.
{}From the commutation relations
$[\alpha_n\,,\,\alpha_m]=n\,\delta_{m+n,0}$
(see \eq{walg0}), it follows that the oscillators
$\alpha_n$ with $n> 0$ ($n < 0$) are destruction (creation)
operators,
and their normal ordering is canonically defined.

The higher spin $\winf$generators can be
constructed in the bosonic language out of the current (\ref{fouri})
through a generalized Sugawara
construction \cite{kac1}. For example, the lowest spin generators
are given by
\barr
V^0(z) &=&\partial \varphi(z)~~~,\nl
V^1(z) &=& {\frac{1}{2}}\ :\left( \partial
\varphi(z)\right)^2 :~~~,\label{bosocur}\\
V^2(z) &=& {\frac{1}{3}}\ :\left( \partial
\varphi(z)\right)^3 : ~~~, \nl
V^3(z) &=& {\frac{1}{4}}\ :\left( \partial \varphi(z)\right)^4
:\,-\, {\frac{3}{20}}\ :\left( \partial^2 \varphi(z)\right)^2 :\,
+\, {\frac{1}{10}} \ :\, \partial \varphi(z) \,\partial^3
\varphi(z)\,
:~~~,
\nonumber
\earr
and their corresponding modes are
\barr
\v^0_n &=& \alpha_n ~~~,\nl
\v^1_n &=& {\frac{1}{2}} \sum_{\ell= -\infty}^{\infty}
:\, \alpha_{\ell}\,\alpha_{n-\ell}\,
:~~~,\label{modeboso}\\
\v^2_n &=& {\frac{1}{3}} \sum_{\ell, k = -\infty}^{\infty}
:\, \alpha_{\ell}\,\alpha_k\, \alpha_{n-\ell-k}\,:~~~,\nl
\v^3_n &=& {\frac{1}{4}} \sum_{\ell,k,m =-\infty}^{\infty}:\,
\alpha_{\ell}\,\alpha_k\, \alpha_m\, \alpha_{n-\ell-k-m}\,:\nl
&&+\ {\frac{1}{20}} \sum_{\ell=-\infty}^{\infty}
(2n-5\ell+1)\,(n-\ell+1)\
:\, \alpha_{\ell}\,\alpha_{n-\ell}\,:~~~.
\nonumber
\earr
As we have stressed above, the effective theory of physically
interesting systems (like the Luttinger model) is a
$W_{1+\infty} \times \wb_{1+\infty}$ conformal theory
with $c={\bar c}=1$.
In the bosonic language, it can be realized by a free compactified
field $\Phi$~\cite{bpz} with euclidean action
\beq
S = \frac{1}{2\pi r} \int d^2 z~\partial \Phi
\,{\overline\partial}\Phi~~~,
\label{bosac}
\eeq
and a compactification radius $r$ defined by $\Phi\ \equiv\ \Phi
+ 2\pi r$.
The equations of motion for $\Phi$ give
$$
\Phi(z,{\overline z})=\varphi(z) + {\overline\varphi}({\overline
z})~~~,
$$
where $\varphi(z)$ and ${\overline\varphi}({\overline z})$ are the
chiral
and antichiral components of the boson field $\Phi$, respectively.
The spectrum of this theory is well-known \cite{bpz} and can be given
in terms of two numbers $m$ and $n$ (with $m \in {\bf Z}$ and
$(m/2-n)
\in {\bf Z}$). In particular, the highest weight states denoted by
$|\,m,n\,\rangle$, are such that
\bea
\v^0_0\,|\,m,n\,\rangle
&=& Q\, |\,m,n\,\rangle  ~~~,\nl
\vb^0_0\,|\,m,n\,\rangle &=& {\overline Q}\,
|\,m,n\,\rangle~~~,
\label{specbo'}
\ena
where the right and left charges $Q$ and ${\overline Q}$ are
respectively
\beq
Q = \frac{m}{2r} + r\,n ~~~~,~~~~{\overline Q} = \frac{m}{2r}
-r\,n~~~.
\label{specbo}
\eeq
Note that the two chiral components of $\Phi$ are not totally
independent since there is an overall constraint of ``charge
conservation''. Furthermore, for general values of $r$,
the charges $Q$ and ${\overline Q}$
are not necessarily integers, as opposed to the fermionic case.
Indeed, it is well-known \cite{bpz} that the free fermion
representation discussed in the previous paragraph is
equivalent to that of a compactified bosonic field
only for $r=1$. However, in the bosonic realization,
one can freely change the value of the compactification radius
without changing the algebra. It is precisely this freedom
which will be exploited in the following sections
to diagonalize the effective hamiltonians of the
Calogero-Sutherland and the Heisenberg models.

\vskip 1.5cm
\section{\bf The $\winf$structure of the Calogero-Sutherland model}
\label{csw}
\bigskip

In this section we demonstrate that the $\winf$algebra is the natural
framework to interpret and understand the results
of \secn{linear}. In particular we show that the effective
hamiltonian of the
Calogero-Sutherland model, \eq{seriesc}, can be nicely written in
terms of $\winf$currents, and that the low-energy spectrum
of the model follows directly from the $\winf$representation theory.
To this aim, we introduce {\it two} sets of $\winf$generators,
one for each Fermi point: the right ones,
denoted by $\v^i_n$, are bilinear forms
in the $a$ oscillators according to \eq{fockw}, whilst the left ones,
denoted by $\vb_n^i$, are bilinear in the $b$ oscillators.
Then, it is easy to see that  Eqs. (\ref{hc0})-(\ref{hc2})
become
\beq
{\cal H}_{(0)} =  \frac{1}{4} (1+g) \left(\v^0_0+
\vb^0_0\right)~~~,
\label{hcv0}
\eeq
\beq
{\cal H}_{(1)} = \left(1+\frac{g}{2}\right) \left(\v^1_0+
\vb^1_0\right)+\frac{g}{2} \sum_{\ell=-\infty}^{\infty}
\v^0_{\ell}\, \vb^0_{\ell}~~~,
\label{hcv1}
\eeq
\bea
{\cal H}_{(2)} &=&
\left(1+\frac{g}{4}\right) \left(\v^2_0+\vb^2_0\right)
-\frac{1}{12} (1+g) \left(\v^0_0+\vb^0_0\right) \nl
&&-\ \frac{g}{4}\sum_{\ell=-\infty}^{\infty} |\ell |
\left( \v^0_{\ell}\,\v^0_{-\ell}+
\vb^0_{-\ell} \,\vb^0_{\ell}+
2\,\v^0_{\ell} \,\vb^0_{\ell}\right)\nl
&&+\ \frac{g}{2}
\sum_{\ell=-\infty}^{\infty} \left( \v^1_{\ell}\,\vb^0_{\ell}+
\v^0_{\ell} \,\vb^1_{\ell} \right) ~~~.
\label{hcv2}
\ena

As explained in \secn{w}, the operators $\v_\ell^0$ and  $\vb_\ell^0$
satisfy a $\u1$ Kac-Moody algebra with central charge $c=1$ (see
\eq{walg0}),
and can be identified with the right and left
modes of a non-chiral bosonic field compactified on a circle,
respectively.
The highest weight states of this Kac-Moody algebra are labeled
by two quantum numbers $\Delta N$ and $\Delta D$, and
will be denoted by $|\Delta N,\Delta D\rangle_0$. The meaning of
these states is particularly clear in the original fermionic
description. In fact, $|\Delta N,\Delta D\rangle_0$
is obtained from the fermionic ground state
$|\Omega \rangle$ by adding $\Delta N$ particles,
and by moving $\Delta D$ particles from the right to the left
Fermi point .
To be precise, given any
two integer numbers $q$ and ${\bar q}$,
we have
$$
|\Delta N , \Delta D \rangle_0 =A_q\, B_{\bar q}
\,|\Omega \rangle~~~,
$$
where
$$
\Delta N =q + {\bar q} ~~~~,~~~~ \Delta D=\frac{q - {\bar q}}{2}~~~,
$$
and
$$
A_{q}=\left\{ \matrix{
a^{\dagger}_1 \,a^{\dagger}_2 \cdots a^{\dagger}_{q}~~~~ &
{\rm for} ~~q=1,2,\ldots~~~~~ \cr
1~~~~~~~~~~~~~ & {\rm for}~~q=0~~~~~~~~~~~~~ \cr
a_0 \,a_{-1}\cdots a_{q+1} & {\rm for}~~q=-1,-2,\ldots\cr}
\right.
$$
and analogously for $B_{{\bar q}}$ with the $b$ oscillators.
The descendant states, denoted by
$|\Delta N , \Delta D ; \{k_i\},\{{\overline k}_j\} \rangle_0$,
have also a simple interpretation in the original fermionic
description: in fact, they
coincide with the
particle-hole excitations obtained from $|\Delta N , \Delta D
\rangle_0$
by acting with operators of the kind $a^{\dagger}_r a_s$,
$b^{\dagger}_{\bar r} b_{\bar s}$ or combinations thereof.

Using the explicit expressions of $\v_0^0$ and $\vb_0^0$ given in
\eq{fockw}, one can easily check that
\bea
\v_0^0 ~|\Delta N , \Delta D ; \{k_i\},\{{\overline k}_j\}
\rangle_0&=&
\left(\frac{\Delta N}{2} \ + \Delta D  \right)
|\Delta N , \Delta D ; \{k_i\},\{{\overline k}_j\} \rangle_0~~~,\nl
\vb_0^0 ~ |\Delta N , \Delta D ; \{k_i\},\{{\overline k}_j\}
\rangle_0 \ &=&
\left(\frac{\Delta N}{2} \ - \Delta D  \right)
|\Delta N , \Delta D ; \{k_i\},\{{\overline k}_j\} \rangle_0~~~.
\label{hwste}
\ena
Comparing these eigenvalues with \eq{specbo}, we deduce that $\v^0_n$
and $\vb_n^0$ are the modes of a
bosonic field compactified on a circle of radius $r_0=1$.
This field describes the density fluctuations of the original
free fermionic fields  of Eqs. (\ref{f+}) and (\ref{f-}).

Let us now consider the $1/N$-term of the effective
hamiltonian given by \eq{hcv1}.
Due to the left-right mixed part proportional to $g$, ${\cal
H}_{(1)}$
is not diagonal on the states
$|\Delta N , \Delta D ; \{k_i\},\{{\overline k}_j\} \rangle_0$
previously considered. However, it is not difficult
to diagonalize it. To do so, we
use the Sugawara construction (see \eq{modeboso}), and replace
$\left(\v_0^1 +\vb_0^1\right)$ with a quadratic form in $\v_\ell^0$
and
 $\vb^0_{\ell}$, so that ${\cal H}_{(1)}$ becomes
\bea
{\cal H}_{(1)} &=& \frac{1}{2}\left(1+\frac{g}{2}\right)
\left[\left(\v_0^0\right)^2+\left(\vb_0^0\right)^2\right]
+\frac{g}{2}\,\v_0^0\,\vb_0^0 \label{hv1}\\
&&+\ \sum_{\ell=1}^\infty
\left[\left(1+\frac{g}{2}\right)\left(
\v_{-\ell}^0\,\v_{\ell}^0+\vb_{-\ell}^0\,\vb_{\ell}^0\right)
+\frac{g}{2}\left(\v_{\ell}^0\,\vb_{\ell}^0+\v_{-\ell}^0\,\vb_{-\ell}^
0\right)
\right]~~~.\nonumber
\ena
This expression exhibits the
essential feature of the algebraic bosonization:
through the Sugawara construction,
a two-fermion term has been replaced with a two-boson term
satisfying the same algebraic properties \cite{matlib}.
The quadratic form in the r.h.s. of \eq{hv1} can now
be diagonalized  by means of the following
Bogoliubov transformation
\barr
\w^0_{\ell}&=&\v^0_{\ell}\ \cosh \beta + \vb^0_{-\ell}\
\sinh \beta ~~~, \nl
\wb^0_{\ell}&=&\v^0_{-\ell}\ \sinh \beta +
\vb^0_{\ell}\ \cosh \beta
\label{bogo}
\earr
for all $\ell$, with
\beq
 \tanh 2\beta =\frac{g}{2+g} ~~~.
\label{angle}
\eeq
In fact, using \eq{bogo} into \eq{hv1},
up to an irrelevant additive constant we get
\beq
{\cal H}_{(1)} = \frac{\lambda}{2}\left[\left(\w_0^0\right)^2
+\left(\wb_0^0\right)^2\right]
+\lambda\,\sum_{\ell=1}^\infty
\left(\w_{-\ell}^0\,\w_{\ell}^0+\wb_{-\ell}^0\,\wb_{\ell}^0\right)~~~,
\label{hw}
\eeq
where
\beq
\lambda \equiv \exp(2\beta)= \sqrt{1+g}~~~.
\label{deflam}
\eeq
In writing \eq{hw} we have used the property that
$\w^0_{\ell}$ and $\wb^0_{\ell}$
satisfy an abelian Kac-Moody algebra with
central charge $c=1$ like the original
operators $\v^0_{\ell}$ and $\vb^0_{\ell}$ (cf. \eq{walg0}).
By means of the generalized
Sugawara construction,
we can then define a new realization of the  $\winf$algebra whose
generators
$\w_n^i$ and $\wb_n^i$ are forms of degree
$(i+1)$ in $\w^0_{\ell}$ and $\wb^0_{\ell}$ respectively, like those
of
\eq{modeboso}.
Consequently, we can rewrite \eq{hw} as follows
\beq
{\cal H}_{(1)} = \lambda\left( \w^1_0 + \wb^1_0 \right)~~~,
\label{h1w}
\eeq
while ${\cal H}_{(0)}$, given in
\eq{hcv0}, simply becomes
\beq
{\cal H}_{(0)} = \frac{\sqrt{\lambda^3}}{4}
\left( \w^0_0+ \wb^0_0 \right)~~~.
\label{h0w}
\eeq
The effective hamiltonian
of the Calogero-Sutherland model up to order $1/N$, {\it i.e.}
\bea
{\cal H}_{(1/N)}&\equiv& \left(2\pi \rho_0\right)^2
\left( {\cal H}_{(0)}+\frac{1}{N}\,{\cal H}_{(1)}\right)
\nl
&=& \left(
2\pi\rho_0  \sqrt{\lambda}\right)^2
\left[\left(\frac{\sqrt{\lambda}}{4}\,\w_0^0
+\frac{1}{N}\,\w_0^1\right)+
\left(\,W~\leftrightarrow~{\overline W}\,\right)
\right]~~~,
\label{h1n}
\ena
exhibits a left-right factorization in the {\it new } realization of
the
$\winf$algebra. In particular, in the
r.h.s. of \eq{h1n} we recognize the typical structure of the
hamiltonian of a conformal
field theory, whose spectrum is known.

Notice that ${\cal H}_{(1/N)}$ is not diagonal on
$|\Delta N , \Delta D ; \{k_i\},\{{\overline k}_j\} \rangle_0$,
because the highest weight states of the
new algebra do not coincide with the
vectors $|\Delta N , \Delta D \rangle_0$, as is clear from \eq{bogo}.
However, since the new charge operators, $\w_0^0$
and $\wb_0^0$, depend only on $\v_0^0$ and $\vb_0^0$,
the Bogoliubov transformation does not mix states
belonging to different Verma moduli.
This implies that the new highest weight vectors are still
characterized by the numbers $\Delta N$ and $\Delta D$ with the
same meaning as before, but their charges are different.
More precisely, the new highest weight states are
defined by
$$
\w^i_\ell ~|\Delta N ; \Delta D \rangle_W =
\wb^i_\ell ~|\Delta N ; \Delta D \rangle_W =0
$$
for all $\ell>0$ and $i\ge  0$, and
\bea
\w_0^0 ~|\Delta N ; \Delta D \rangle_W  &=&
\left(\sqrt{\lambda}\,\frac{\Delta N}{2}+
\frac{\Delta D}{\sqrt{\lambda}}\right)
|\Delta N ; \Delta D \rangle_W \nl
\wb_0^0 ~|\Delta N ; \Delta D \rangle_W  &=&
\left(\sqrt{\lambda}\,\frac{\Delta N}{2}-
\frac{\Delta D}{\sqrt{\lambda}}\right)
|\Delta N ; \Delta D \rangle_W~~~.
\label{vdo}
\ena
Comparing these last two equations with \eq{specbo},
we can deduce that $\w^0_\ell$
and $\wb_\ell^0$ are the modes of a
bosonic field compactified on a circle of radius $r=1/\sqrt{\lambda}=
\exp(-\beta)$. This field describes the density fluctuations
of the {\it interacting} fermions of the Calogero-Sutherland
model.

The highest weight states $|\Delta N , \Delta D \rangle_W$
together with their descendants, denoted by
$|\Delta N , \Delta D ; \{k_i\},\{{\overline k}_j\} \rangle_W$,
form a {\it new} bosonic basis for our
theory that has no simple expression in
terms of the original free fermionic degrees of freedom. In fact,
as is well-known, the Bogoliubov transformation,
\eq{bogo}, is non-local in the fermionic operators.
The main property of this new basis is that it diagonalizes
the effective hamiltonian of the Calogero-Sutherland model
up to order $1/N$. In fact, using Eqs. (\ref{wcs}),
(\ref{h1n}) and (\ref{vdo}),
it is easy to check that
\beq
{\cal H}_{(1/N)}
{}~|\Delta N , \Delta D ; \{k_i\},\{{\overline k}_j\} \rangle_W =
{\cal E}_{(1/N)}
{}~|\Delta N , \Delta D ; \{k_i\},\{{\overline k}_j\} \rangle_W
\label{eig}
\eeq
where
\beq
{\cal E}_{(1/N)}= \left(
2\pi\rho_0  \sqrt{\lambda}\right)^2\Bigg[\frac{\lambda}{4}\,\Delta N
+
\frac{1}{N}\left(\lambda\,\frac{\left(\Delta N\right)^2}{4}  +
\frac{\left(\Delta D\right)^2}{\lambda} +
k+{\overline k} \right)\Bigg] ~~~,
\label{eigen}
\eeq
with $k=\sum\limits_i k_i$ and ${\overline k}=\sum\limits_j
{\overline k}_j~$.
These eigenvalues are clearly degenerate when $k \geq 2$ or
${\overline k} \geq 2$.
Notice that \eq{eigen} can be written also as follows
$$
{\cal E}_{(1/N)}= \mu \,\Delta N + \frac{2\pi v}{L}
\left(\lambda\,\frac{\left(\Delta N\right)^2}{4}  +
\frac{\left(\Delta D\right)^2}{\lambda} +
k+{\overline k} \right) ~~~,
$$
where $\mu$ is the chemical potential and $v$ the Fermi velocity.
Examining the structure of the energy eigenvalues (\ref{eigen})
and comparing with those at $g=0$,
one can say that, up to order $1/N$,
the Calogero-Sutherland interaction induces
the following three effects:
\begin{enumerate}
\item{a rescaling of the chemical potential
\beq
\mu_0=\frac{\left(2\pi\rho_0  \right)^2}{4}
{}~\longrightarrow~
\mu= \lambda^2\,\mu_0~~~,
\label{chempot}
\eeq}
\item{a rescaling of the Fermi velocity of the particles
\beq
v_0=2\pi\rho_0
{}~\longrightarrow~
v= \lambda \,v_0~~~,
\label{fervel}
\eeq}
\item{a change in the compactification
radius of the bosonic field describing the fermion
density fluctuations
\beq
r_0=1~\longrightarrow~
r=\frac{r_0}{\sqrt{\lambda}}~~~.
\label{comprad}
\eeq}
\end{enumerate}

It is interesting to observe that
the rescalings in Eqs. (\ref{chempot})-(\ref{comprad})
have their origin in the backward scattering processes of the
Calogero-Sutherland model, which, as we have remarked at
the end of \secn{linear}, are
the only interactions that contribute to the effective
hamiltonian to order $1/N$.
In particular, to lowest order in $g$,
the change in the compactification
radius is induced
by the left-right mixed terms of
${\cal H}_{(1)}$. These are indeed
the $1/N$-terms of the backscattering hamitonian (\ref{ht}), written
in the bosonized language.
Such terms have the generic form $\v^0_\ell\,\vb^0_\ell\,$,
and have conformal
dimension $(1,1)$. Therefore, they are marginal operators, which
cannot destroy the conformal symmetry of the free
theory, but only change the realization of the conformal algebra
\cite{bpz}.
In fact, these terms can be regarded as a marginal perturbation
to \eq{bosac}
which drives the theory out from the free realization in terms of
$\v_\ell^0$ and $\vb_\ell^0$ to the interacting realization in terms
of $\w_\ell^0$ and $\wb_\ell^0$. In this flow, the central charge of
the conformal algebra remains unchanged while the compactification
radius of the bosonic field varies according to \eq{comprad}.

The rescalings of the chemical potential and the Fermi velocity,
Eqs. (\ref{chempot}) and (\ref{fervel}), have instead a different
interpretation. In fact, to lowest order in $g$, they are produced
by the left and right diagonal terms proportional to
$g$ in ${\cal H}_{(0)}$ and ${\cal H}_{(1)}$. If we trace back their
origin, we see that these terms arise from the two-body part of
the backscattering hamiltonian (\ref{hdos}). Therefore, they are a
normal ordering effect.

It is remarkable that despite their different origins,
the rescalings in Eqs. (\ref{chempot})-(\ref{comprad})
are characterized by only {\it one} function of the coupling
constant, namely the parameter $\lambda$ defined in \eq{deflam}.
This fact implies that they
are not independent from one another; for example, one has
\beq
v_0\,r_0^2 = v\,r^2~~~.
\label{vr}
\eeq
This relation is typical of the Luttinger model \cite{solyo},
and actually
holds true for all systems whose hamiltonian at order $1/N$ has the
same form as in Eqs. (\ref{hv1}) or (\ref{h1w}), that is, in all cases
for which the interaction can be simply taken into
account by means of a Bogoliubov transformation like \eq{bogo}.
However, not all models fit into this category.

At this point a few comments are in order. We should keep in
mind that the derivation of the effective theory, as presented
in \secn{linear}, is strictly perturbative; thus,
in all previous formulas, we should always understand
a perturbative expansion in the coupling constant
$g$, and keep only the first order corrections.
However, if we limit our analysis to the $1/N$-terms,
nothing prevents us from improving our results
and extend them to all orders in $g$. Indeed,
when we perform the Bogoliubov transformation
(\ref{bogo}), we diagonalize the hamiltonian
${\cal H}_{(1)}$ {\it exactly}, and the resulting
expression depends on the coupling
constant only through $\lambda$, which contains
all powers of $g$ (see \eq{deflam})! This improvement is
a well-known fact in the Luttinger model \cite{solyo},
but we would like to stress that in our case
it can be done only if we disregard
the $O(1/N^2)$-terms of the hamiltonian. In fact,
as we shall see momentarily, the Bogoliubov transformation
(\ref{bogo}) does not diagonalize ${\cal H}_{(2)}$.

To investigate this issue, let us
analyze the $1/N^2$-term of the effective
hamiltonian given by
\eq{hcv2}. Using the generalized Sugawara
construction (\ref{modeboso}), we first rewrite
${\cal H}_{(2)}$ as a cubic form in $\v^0_{\ell}$
and $\vb^0_{\ell}$, and then perform
the Bogoliubov transformation (\ref{bogo}) in order
to express it in terms of the new generators of the $\winf$algebra.
A straightforward calculation leads to
$$
{\cal H}_{(2)} = {\cal H}_{(2)}' \ +\ {\cal H}_{(2)}''
$$
where
\bea
{\cal H}_{(2)}' &=&
\sqrt{\lambda} \left( \w^2_0+\wb^2_0 \right)-
\frac{\sqrt{\lambda^3}}{12}\left(\w_0^0+\wb_0^0\right)
\nl
&&-\ \frac{g}{2 \lambda}
\,\sum_{\ell=1}^{\infty}\,\ell\left(
\w^0_{-\ell}\,\w^0_{\ell}+ \wb^0_{-\ell}\,\wb^0_{\ell}
\right) ~~~,
\label{h2w'}
\ena
and
\beq
{\cal H}_{(2)}'' \,=\, -\,\frac{g}{2 \lambda}
\,\sum_{\ell=1}^{\infty}\,\ell\left(
\w^0_{\ell}\,\wb^0_{\ell}+ \w^0_{-\ell}\,\wb^0_{-\ell}
\right)~~~.
\label{h2w''}
\eeq
Neither ${\cal H}_{(2)}'$ nor
${\cal H}_{(2)}''$ are diagonal in the basis
$|\Delta N , \Delta D ; \{k_i\},\{{\overline k}_j\} \rangle_W$
considered so far. In general, these states are
not eigenstates of $\left(\w_0^2+\wb_0^2\right)$, and hence
cannot be eigenstates of
${\cal H}_{(2)}'$; moreover, since they have definite
values of $k$ and $\bar k$, they cannot be eigenstates
of ${\cal H}_{(2)}''$ either, because this operator
mixes the left and right sectors.

It is not difficult, however, to overcome
these problems. Since
${\cal H}_{(2)}'$ and ${\cal H}_{(1/N)}$ commute with each
other, it is always possible to find suitable
combinations of the states
$|\Delta N , \Delta D ; \{k_i\},\{{\overline k}_j\} \rangle_W$
with fixed $k$ and ${\overline k}$ that diagonalize
simultaneously ${\cal H}_{(2)}'$ and ${\cal H}_{(1/N)}$
(see below for a few explicit examples).
Notice that by diagonalizing also
${\cal H}_{(2)}'$, we lift the degeneracy of the
spectrum that appeared to order $1/N$.
We denote the space of the eigenstates of
${\cal H}_{(2)}'$ and ${\cal H}_{(1/N)}$ by
${\cal W}\,(\Delta N , \Delta D)$.

The term ${\cal H}_{(2)}''$ instead can be treated perturbatively,
but only to first order in $g$; in fact
at higher orders, also the spurious states
introduced when sending the momentum cutoff $\Lambda_0 \to \infty$
(see the discussion before \eq{f+} in \secn{linear}) would contribute
as intermediate states.
These contributions, however,
would be meaningless because the hamiltonian to order $O(1/N^2)$ is
not even bounded below.
{}From \eq{h2w''} it is easy to check
that ${\cal H}_{(2)}''$ has vanishing
expectation value on any state belonging
to ${\cal W}(\Delta N , \Delta D)$, and
thus, according to (non-degenerate) perturbation theory,
${\cal H}_{(2)}''$ has no effect on the energy spectrum
to first order in $g$.

In view of these considerations, we can neglect
${\cal H}_{(2)}''$ and regard as the effective
hamiltonian of the Calogero-Sutherland model
the following operator
\bea
{\cal H}_{CS} &\equiv&
{\cal H}_{(1/N)} + \left(
2\pi\rho_0 \right)^2
\frac{1}{N^2}\,{\cal H}_{(2)}'\nl
&=&\left(
2\pi\rho_0  \sqrt{\lambda}\right)^2
\left\{\left[\frac{\sqrt{\lambda}}{4}\,\w_0^0
+\frac{1}{N}\,\w_0^1+
\frac{1}{N^2}\left(\frac{1}{\sqrt{\lambda}}\,\w_0^2
-\frac{\sqrt{\lambda}}{12}\,\w_0^0 \right.\right.
\right.\nl
&&-\ \left.\left.\left.
\frac{g}{2\lambda^2}\,\sum_{\ell=1}^\infty
\,\ell~\w_{-\ell}^0\,\w_\ell^0\right)
\right]+\left(\,W~\leftrightarrow~{\overline W}\,\right)
\right\}~~~,
\label{hcsf}
\ena
where $\lambda$ is defined
in \eq{deflam}. Obviously, to be consistent with
our perturbative approach, in the r.h.s.
of \eq{hcsf} we should keep
only terms that are linear in $g$.

It is now interesting to
compare the eigenvalues of ${\cal H}_{CS}$
with the exact low-energy spectrum of the Calogero-Sutherland model
obtained from the Bethe Ansatz solution \cite{sut,kaya}.
It is known that the energy (with respect to the ground state) of any
configuration of the system can be written as
\beq
\tilde{\cal E} = \sum_{j=1}^N p_j^2 - E_0
\label{ener}
\eeq
where $E_0$ is the ground state energy and $p_j$ are the
pseudomomenta
of the particles which satisfy the Bethe Ansatz equations. In our
case
these take a particularly simple form;
in fact they are
\beq
L \, p_j = 2\pi\, I_j + \pi \left( \xi-1 \right)
\sum_{l=1}^N {\rm sgn} \left(p_j-p_\ell \right) ~~~,
\label{pseudo}
\eeq
where
\beq
\xi=\frac{1+\sqrt{1+2g}}{2}~~~,
\label{xi}
\eeq
and $I_j$ are the integer quantum numbers that specify the levels
occupied by the particles. For example, the ground state is
characterized by the set
$$
\left\{ I_j^0\right\} \ =\ \{-n_F,\, -n_F+1,\, \dots \, , n_F \} ~~~.
$$
with $n_F=(N-1)/2$~.

A low-lying excitation above the ground state can be obtained
in three different ways:
by adding $\Delta N$ particles to the system ($\Delta N \in {\bf
Z}$),
by moving $\Delta D$ particles from the left to the right Fermi point
($\Delta N /2-\Delta D \in {\bf Z}$),
or by creating particle-hole pairs at levels $n_j$ and ${\overline
n_j}$
on the right and on the left respectively.
Therefore, a generic low-lying excitation is labeled by the
following quantum numbers
\beq
I_j\ =\ {\tilde I}_j^0 + \Delta D - {\overline n}_j +  n_{N'-j+1}
\label{Ii}
\eeq
where $j=1,2, \dots, N'$, and
$$
\left\{ {\tilde I}_j^0\right\} \ =\ \left\{-\frac{N'-1}{2},\,
-\frac{N'-1}{2}+1,\, \dots, \,\frac{N'-1}{2}\right\}
$$
with
$$
N'= N + \Delta N ~~~.
$$
Moreover, the integer numbers $n_j$ are ordered
according to
$$
n_1\geq n_2\geq \dots \geq 0
$$
and are different from zero only if $j << N$ (and analogously for the
${\overline n}_j$).

By using Eqs. (\ref{ener}) and (\ref{pseudo}) and generalizing to
order $1/N^2$ the procedure presented in Ref.~\cite{kaya}, we can
easily derive the exact energy of the excitation described by the
numbers
(\ref{Ii}); this is
\bea
\tilde{\cal E}&=&\left(
2\pi\rho_0  \sqrt{\xi}\right)^2
\left\{\left[\frac{\sqrt{\xi}}{4}\,Q+\frac{1}{N}
\left(\frac{1}{2}\,Q^2+n\right)
+\frac{1}{N^2}\left(\frac{1}{3\sqrt{\xi}}\,Q^3
-\frac{\sqrt{\xi}}{12}\,Q\right.\right.\right.
\label{eba} \\
&&+\ \left.\left.
\frac{2n}{\sqrt{\xi}}\,Q + \frac{\sum_j n_j^2}{\xi}
-\sum_j \left(2j-1\right) n_j\right)\right]
+\left(Q\ \leftrightarrow \ {\overline Q}~,
{}~\{n_j\} \ \leftrightarrow \ \{{\overline n}_j\} \right)\Bigg\}~~~,
\nonumber
\ena
where
$$ n \ =\ \sum_j n_j ~~~~,~~~~
{\overline n} \ =\ \sum_j {\overline n}_j
$$
and
\beq
Q=\sqrt{\xi}\,\frac{\Delta N}{2}+
\frac{\Delta D}{\sqrt{\xi}}~~~~,~~~~
{\overline Q}=\sqrt{\xi}\,\frac{\Delta N}{2}-
\frac{\Delta D}{\sqrt{\xi}}~~~,
\label{Q}
\eeq
Of course, being an exact
result, \eq{eba} holds to all orders in $g$.
Comparing Eqs. (\ref{deflam}) and (\ref{xi}), we
see that
\beq
\xi=\lambda+O(g^2)~~~.
\label{xilam}
\eeq
Thus, to first order in $g$,
$Q$ and ${\overline Q}$ of \eq{Q}
coincide with the eigenvalues of $\w_0^0$ and
$\wb_0^0$ given in \eq{vdo}; conversely,
these latter can be interpreted as the first-order
approximation to the exact ones.
Notice that $Q$ and ${\overline Q}$ have
the structure of the zero mode charges of a non-chiral
bosonic field compactified on a circle of radius
\beq
{\tilde r} = \frac{1}{\sqrt{\xi}}~~~.
\label{rtil}
\eeq
Indeed, this is the exact value of the compactification
radius of the bosonic field describing the density fluctuations
of the fermions in the Calogero-Sutherland model \cite{kaya}.
{}From \eq{eba} we can also see that the exact value of the
chemical potential is
\beq
{\tilde \mu} = \frac{ \left(2\pi\rho_0\right)^2}{4} \,\xi^2~~~,
\label{xit}
\eeq
while the exact Fermi velocity is
\beq
{\tilde v} = 2\pi\rho_0\,\xi~~~.
\label{vti}
\eeq
These expressions are similar to those in
Eqs. (\ref{chempot})-(\ref{comprad}) with $\xi$ in place of
$\lambda$.
Of course, due to \eq{xilam},
${\tilde r}$, ${\tilde \mu}$ and ${\tilde v}$ coincide, respectively,
with $r$, $\mu$ and $v$, to first order in $g$.
It is worthwhile pointing out that all
low-energy effects of the Calogero-Sutherland
interaction are encoded entirely in a unique quantity, namely
the parameter $\xi$,
which in the Bethe Ansatz literature is known as
dressed charge factor \cite{kore}.

Since the exact results
can be obtained from the perturbative ones simply by changing
$\lambda$ into $\xi$, we are led to conjecture that
the {\it exact} effective hamiltonian of the Calogero-Sutherland
model is given by \eq{hcsf} with $\xi$, defined in \eq{xi},
in place of $\lambda$, that is
\bea
\tilde{\cal H}_{CS} &=&\left(
2\pi\rho_0  \sqrt{\xi}\right)^2
\left\{\left[\frac{\sqrt{\xi}}{4}\,\w_0^0
+\frac{1}{N}\,\w_0^1+
\frac{1}{N^2}\left(\frac{1}{\sqrt{\xi}}\,\w_0^2
-\frac{\sqrt{\xi}}{12}\,\w_0^0 \right.\right.
\right.\nl
&&-\ \left.\left.\left.
\frac{g}{2\xi^2}\,\sum_{\ell=1}^\infty
\,\ell~\w_{-\ell}^0\,\w_\ell^0\right)
\right]+\left(\,W~\leftrightarrow~{\overline W}\,\right)
\right\}~~~.
\label{hcsf1}
\ena
We may consider this operator as a non-perturbative improvement
of ${\cal H}_{CS}$ which was derived in perturbation theory.

Evidence for the validity of our conjecture, which is certanly true
to
order $1/N$ (see \cite{kaya}), is provided
by the calculation of the eigenvalues of ${\tilde {\cal H}}_{CS}$.
We will check on same explicit examples that these eigenvalues
coincide
with the exact energy of the low-lying excitations given in \eq{eba}.
To this aim, let us first consider the highest weight state
$|\Delta N , \Delta D \rangle_W$ that satisfies
\eq{vdo} with $\xi$ in place of $\lambda$.
Using Eqs. (\ref{weig}) and (\ref{wcs}) for $c=1$,
it is immediate to see that
$|\Delta N , \Delta D \rangle_W$ is an eigenstate of
${\tilde {\cal H}}_{CS}$ whose energy is given by \eq{eba} with
$\{n_j\}=\{{\overline n}_j\}=\{0,0,\dots\}$.
Thus, $|\Delta N , \Delta D \rangle_W$
is the state that describes a low-lying excitation without
particle-hole
pairs.

Let us now consider the state
\beq
\w_{-1}^0\,|\Delta N , \Delta D \rangle_W  ~~~.
\label{1,0}
\eeq
Using again \eq{wcs}, we can see that this is an eigenstate of
${\tilde {\cal H}}_{CS}$ with energy given by \eq{eba} with
$\{n_j\}=\{1,0,\dots\}$ and $\{{\overline n}_j\}=\{0,0,\dots\}$.
Thus (\ref{1,0}) is a state with a right particle-hole pair at
level $1$. Similarly, the state
\beq
\w_{-1}^0\,\wb_{-1}^0\,|\Delta N , \Delta D \rangle_W
\label{1,1}
\eeq
is an eigenstate of ${\tilde {\cal H}}_{CS}$ with energy given
by \eq{eba} with $\{n_j\}=\{{\overline n}_j\}=\{1,0,\dots\}$
and represents a state with one particle-hole pair at level $1$ on
the right and one particle-hole pair at level $1$ on
the left.

These calculations can be simply generalized to higher levels, where
the structure of the states is less trivial.
For example, at level $(2,0)$ there are two eigenstates for
${\tilde {\cal H}}_{CS}$:
\bea
&& \left(\w_{-2}^0+\sqrt{\xi}\,\w_{-1}^0\,\w_{-1}^0\right)
|\Delta N , \Delta D \rangle_W ~~~,
\label{2,0} \\
&& \left(\w_{-2}^0-\frac{1}{\sqrt{\xi}}\,\w_{-1}^0\,\w_{-1}^0\right)
|\Delta N , \Delta D \rangle_W ~~~,
\label{2,0*}
\ena
It is easy to see that these two states
have the same energy up to order $1/N$, but actually
are not degenerate due to the $1/N^2$-term of the effective
hamiltonian
${\tilde {\cal H}}_{CS}$. The energy of (\ref{2,0}) turns out to be  
given
exactly by \eq{eba} with $\{n_j\}=\{2,0,\dots \}$ and
$\{{\overline n}_j\}=\{0,0,\dots \}$, and therefore the state  
(\ref{2,0})
can be associated to a single particle-hole excitation of level $2$
on the right.
The energy of (\ref{2,0*}) is, instead, given by \eq{eba} with
$\{n_j\}=\{1,1,0,\dots,0\}$ and $\{{\overline n}_j\}=\{0,0,\dots,0\}$
and coincides with the Bethe Ansatz energy for two particle-hole
excitations at level $1$ on the right.

Increasing the level, one gets more states (in general at level
$(n,{\overline n})$ we have $g(n) \times g({\overline n})$ states,
where $g(n)$ is the number of partitions of $n$).
For instance, at level $(3,0)$ there are three eigenstates
of ${\tilde{\cal H}}_{CS}$, which are
\bea
&&\left(\w_{-3}^0+
\frac{3\sqrt{\xi}}{2}\,\w_{-2}^0\,\w_{-1}^0
+\frac{\xi}{2}\,\w_{-1}^0\,\w_{-1}^0\,\w_{-1}^0\right)
|\Delta N , \Delta D \rangle_W~~~,
\label{3,0} \\
&&\left(\w_{-3}^0+ \left(\sqrt{\xi}-\frac{1}{\sqrt{\xi}}\right)
\w_{-2}^0\,\w_{-1}^0
-\,\w_{-1}^0\,\w_{-1}^0\,\w_{-1}^0\right)
|\Delta N , \Delta D \rangle_W~~~,
\label{3,0*} \\
&&\left(\w_{-3}^0 -
\frac{3}{2\sqrt{\xi}}\,\w_{-2}^0\,\w_{-1}^0
+\frac{1}{2\xi}\,\w_{-1}^0\,\w_{-1}^0\,\w_{-1}^0\right)
|\Delta N , \Delta D \rangle_W~~~,
\label{3,0**}
\ena
and their energies are given by
\eq{eba} with
\bea
\{n_j\}=\{3,0,0,0,\dots \} ~&,&~\{{\overline n}_j\}=\{0,0,\dots\}~~~,
\nl
\{n_j\}=\{2,1,0,0,\dots \} ~&,&~\{{\overline n}_j\}=\{0,0,\dots\}~~~,
\nl
\{n_j\}=\{1,1,1,0,\dots \} ~&,&~\{{\overline n}_j\}=\{0,0,\dots\}~~~,
\nonumber
\ena
respectively.

These examples clearly support our conjecture, and show that the
eigenstates of ${\tilde{\cal H}}_{CS}$ are in one to one
correspondence with the states of the conformal Verma module
at a given level.
When one considers
the subleading $1/N^2$ terms in addition to the leading
(conformal) $1/N$-term, formulating
the model in the context of the $\winf${\it extended} conformal
field theories, the degeneracy of the Verma module at a given level
is completely removed for generic values of $\xi$ \footnote{
The degeneracy of the energy under the exchange $Q \,\leftrightarrow  
\,
{\overline Q}$ and $\{n_j\} \, \leftrightarrow \, \{{\overline  
n}_j\}$
obviously remains.}.
This piece of information is clearly important to compute the exact
partition and correlation functions in the low-energy regime.

\vskip 1.5cm
\section{\bf The $\winf$structure
of the Heisenberg model in a magnetic field}
\label{heisw}
\bigskip

The main issue of this section will be to rewrite the hamiltonian of  
the
Heisenberg model in terms of $\winf$generators as we have done in the
Calogero-Sutherland case. However,
it is first necessary to establish the
relationship between the magnetization $\sigma$
and the external field $B$.
This can be easily obtained by requiring that the energy of the
excitations vanishes on the Fermi surface, {\it i.e.} by equating
to zero
the coefficient of the operator $\sum\limits_{r=-\infty}^\infty
\left(:a_r^\dagger\,a_r:+:b_r^\dagger\,b_r:\right)$
in the hamiltonian (\ref{series}).
To the leading order in the $1/N$
expansion, this requirement turns into the condition
\beq
-\,B + J_z\sigma + \sin\frac{\pi\sigma}{2} +
\frac{J_z}{\pi}\sin\pi\sigma = 0 ~~~.
\label{bsigma}
\eeq
This equation defines the magnetization in terms of the magnetic
field, and therefore fixes the position of the Fermi points $\pm n_F$
according to \eq{nf}.
It also implies that our approach is meaningful only for
$|B|< B_c$, where the critical field
$$
B_c \ = \ 1 + J_z
$$
corresponds to magnetization $\sigma = 1$.
\eq{bsigma} can be explicitly solved when $B \to 0$
and $B \to B_c$. In the first case one has
\beq
\sigma \ =\ \frac{2}{\pi} \left(1+\frac{4J_z}{\pi}\right)^{-1} B~~~,
\label{si0}
\eeq
while in the second case one obtains
\beq
\sigma\ =\ 1 - \frac{2}{\pi}\sqrt{2\left(B_c-B\right)}~~~.
\label{sic}
\eeq
It is easy to check that these values agree, to
first order in $J_z$, with the exact magnetization
derived from the Bethe Ansatz solution
of the Heisenberg model (see for instance \cite{kore}).
For this agreement to occur, the
last term in \eq{bsigma} is crucial.
This term is produced by the two-body part
of the backscattering hamiltonian, \eq{hibac2'}, and thus is a normal
ordering effect.

We are now in the position of writing the effective hamiltonian of  
the
Heisenberg model (\ref{series}) in terms of the $\winf$generators.  
Like
in the Calogero-Surtherland case, here also we introduce two sets of
$\winf$currents, $\v_\ell^i$ and $\vb_\ell^i$, represented  
respectively
as bilinear fermionic forms in the $a$ and $b$ oscillators according  
to
\eq{fockw}. Then, using \eq{bsigma}, the hamiltonian (\ref{series})
becomes
\beq
{\cal H} =
\sum_{k=1}^{\infty} \left( \frac{2\pi}{N} \right)^k {\cal  
H}_{(k)}~~~,
\label{hamd}
\eeq
where the first terms are
\barr
{\cal H}_{(1)} &=& \left(\cos\frac{\pi\sigma}{2} +
\frac{J_z}{\pi}\cos\pi\sigma \right)
\left(\v^1_0+ \vb^1_0\right)+
 \frac{J_z}{2\pi} \sum_{\ell=-\infty}^{\infty}
\left( \v^0_{-\ell}\,\v^0_{\ell}+
\vb^0_{-\ell}\, \vb^0_{\ell}\right) \nl
&&+\ \frac{J_z}{\pi} \left(\cos\pi\sigma + 1\right)
\sum_{\ell=-\infty}^{\infty}\v^0_{\ell}\,\vb^0_{\ell}~~~,
\label{hwh1}
\earr
\barr
{\cal H}_{(2)} &=& -\ \frac{1}{2}\left(\sin\frac{\pi\sigma}{2} +
\frac{J_z}{\pi}\sin\pi\sigma \right) \left[
\left(\v^2_0 + \vb^2_0\right) - \frac{1}{12}
\left(\v^0_0+\vb^0_0\right)\right] \nl
&&-\ \frac{J_z}{\pi} \sin \pi\sigma
\sum_{\ell=-\infty}^{\infty} \left( \v^1_{\ell}\, \vb^0_{\ell}
+\v^0_{\ell}\,\vb^1_{\ell}\right)
\label{hwh2}
\earr
\barr
{\cal H}_{(3)} &=&-\ \frac{1}{6} \left(
\cos\frac{\pi\sigma}{2} +\frac{J_z}{\pi}\cos\pi\sigma \right)
\left(\v^3_0+\vb^3_0\right)\nl
&&+\ \frac{1}{10} \left(\frac{7}{12}\cos\frac{\pi\sigma}{2} +
\frac{J_z}{\pi}\cos\pi\sigma \right)
\left(\v^1_0+ \vb^1_0\right) \nl
&&-\ \frac{J_z}{4\pi} \sum_{\ell=-\infty}^{\infty}
\left[\,{\ell}^2 \left( \v^0_{-\ell}\,\v^0_{\ell}
+ \vb^0_{-\ell}\,\vb^0_{\ell}\right)+
\left(2{\ell}^2 +\frac{1}{3}({\ell}^2-1)\cos\pi\sigma \right)
\v^0_{\ell}\,\vb^0_{\ell}\right] \nl
&&-\ \frac{J_z}{2\pi} \cos\pi\sigma
\sum_{\ell=-\infty}^{\infty}\left(\v^2_{\ell}\, \vb^0_{\ell}
+ 2 \v^1_{\ell} \,\vb^1_{\ell}+ \v^0_{\ell}\,\vb^2_{\ell}
\right)~~~.
\label{hwh3}
\earr
We remark that ${\cal H}_{(0)}$
vanishes due to \eq{bsigma};
furthermore after inclusion of the $O(1/N^2)$-corrections to
\eq{bsigma}, also the second line of \eq{h2f} vanishes.

These equations display the effective hamiltonian of the Heisenberg
model as a combination of $\winf$currents. However,
it is important to recall that in the absence of an external magnetic
field, Umklapp terms should be also taken into account; these terms
(see \eq{hi}) would give contributions to ${\cal H}$ of the
form $a^\dagger_.\,b_.\, a^\dagger_.\,b_.$ or
$b^\dagger_.\,a_.\, b^\dagger_.\,a_.$, which destroy conformal
invariance and cannot be written in terms
of $\winf$generators. This fact should not come as a surprise
because it is well-known that Umklapp terms spoil charge-current
conservation, which is expressed by the $\winf$algebra.

We now focus on the $1/N$-term of the effective hamiltonian
given by \eq{hwh1}, and proceed exactly as
in the Calogero-Sutherland case. We first replace $\left(\v_0^1 +
\vb_0^1\right)$ with a quadratic form in $\v_\ell^0$ and $\vb_\ell^0$
by means of the Sugawara construction, and then introduce new
operators $\w_\ell^0$ and $\wb_\ell^0$ according to \eq{bogo}.
The resulting quadratic form becomes diagonal in the new
generators if, to first order in $J_z$, we choose
\footnote{Of course the diagonalization of ${\cal H}_{(1)}$ can be  
done
exactly to all orders in $J_z$; however, as it will be clear in the
following, only the first order in $J_z$ is meaningful.}
\beq
\tanh 2\beta = \frac{2J_z}{\pi} \cos \frac{\pi\sigma}{2}~~~.
\label{an1}
\eeq
This choice implies also that $\w^0_\ell$ and $\wb^0_\ell$ are the
modes of a bosonic field compactified on a circle of radius
\beq
r=\exp(- \beta) \simeq
1 - \frac{J_z}{\pi}\cos\frac{\pi\sigma}{2} ~~~.
\label{radius}
\eeq
If we use once more the Sugawara construction, we finally obtain
\beq
{\cal H}_{(1/N)} \equiv \frac{2\pi}{N} {\cal H}_{(1)} =
\frac{2\pi}{N} \left[ \cos\frac{\pi\sigma}{2}
+ \frac{J_z}{\pi} \left( \cos\pi\sigma + 1\right)\right]
\left( W^1_0\ +\ \wb^1_0 \right) ~~~.
\label{anhb}
\eeq
The r.h.s. clearly exhibts the well-known fact that the
effective hamiltonian of the Heisenberg model to order $1/N$
is that of a $c=1$ conformal field theory.
{}From \eq{anhb}, we also read that the Fermi
velocity of the low-energy excitations is
\beq
v = \cos\frac{\pi\sigma}{2} + \frac{J_z}{\pi}
\left( \cos\pi\sigma + 1\right)
\label{vf}
\eeq
Obviously both the Fermi velocity $v$ and the compactification radius  
$r$
depend on $B$ through $\sigma$, and whenever \eq{bsigma} can be  
solved,
one can obtain their explicit relation with the magnetic field.
In particular, when $B \to 0$ from \eq{si0} we have
\barr
r &=& \left(1-\frac{J_z}{\pi}\right)\left(1+ \frac{J_z}{2\pi}
B^2\right) ~~~,\nl
v &=& 1 + \frac{2J_z}{\pi}~~~,
\label{b0}
\earr
while, when $B \to B_c$  from \eq{sic} we have
\barr
r &=& 1 - \frac{J_z}{\pi}\sqrt{2\left(B_c-B\right)}~~~, \nl
v &=& \sqrt{2\left(B_c - B\right)}~~~.
\label{bbc}
\earr
We observe that both $v$ and $r$ have the right
asymptotic behavior when $B$ is close either to the critical field or  
to
zero, coinciding to first order in $J_z$ with the result
obtained by Bethe Ansatz (see Ref.~\cite{kore}).

It is interesting to realize that the changes of
$r$ and $v$ induced by the Bogoliubov transformation
from $\{\v^0_\ell,\vb^0_\ell\}$ to $\{\w^0_\ell,\wb^0_\ell\}$
have two different origins. In fact, the compactification
radius changes from $r_0=1$ to the value $r$ given in \eq{radius},
as a consequence of the left-right mixing term
proportional to $\v^0_\ell\,\vb^0_\ell$ in \eq{hwh1}.
This term is a marginal operator which
originates both from the backward and the forward scattering  
hamiltonians
to order $1/N$ (see Eqs. (\ref{hibac4'}) and (\ref{hifor4'})).
This is to be contrasted with the situation of the  
Calogero-Sutherland model,
where only the backscattering part of the hamiltonian
contributes at order $1/N$, since
the forward scattering terms are
$O(1/N^2)$ (see \eq{hif1cs}).

The change in the Fermi velocity from  
$v_0=\cos\left(\pi\sigma/2\right)$
to $v$ given in
\eq{vf} is, instead, due to the left and right diagonal terms
proportional to $J_z$ in ${\cal H}_{(1)}$. These terms
have two distinct sources: one is the diagonal part of the forward
scattering hamiltonian (\ref{hifor4'}), and the other is the
two-body part of the backward scattering hamiltonian (\ref{hibac2'}).
The latter is clearly a normal ordering effect.
We would like to stress that only after taking into account
both kinds of terms the Fermi velocity $v$ agrees with the
value given by the Bethe Ansatz solution to first order in $J_z$.
Notice also that
\beq
v_0\,r_0^2 = v\,r^2~~~.
\label{vr'}
\eeq
This relation is typical of the Luttinger systems, as we
mentioned in \secn{csw}, but it is not a property of the
exact Bethe Ansatz solution of the Heisenberg model.
In fact, in this case \eq{vr'} holds only to first order in $J_z$  
(see
Ref.~\cite{kore}).
This strongly suggests that, in order to have complete agreement  
between
the exact solution and ours, we would need at least
two different scaling functions:
one for the Fermi velocity and one for the compactification
radius. However, the Luttinger model approach can provide only one.
Thus, from now on we will limit our considerations to the first
perturbative order in $J_z$, where this problem does not exist.
This fact was already observed in Ref.~\cite{lupe}. However,
due to a different normal ordering prescription, only the
compactification radius was found to be consistent with the Bethe
Ansatz solution to first order in $J_z$. On the contrary, the Fermi
velocity ($c$ in the notation of Ref.~\cite{lupe}) turned out to be
different from the exact value, even to first order in $J_z$.
The reason for this is that the backscattering
hamiltonian of Ref.~\cite{lupe} did not require a rearrangement
of the fermionic oscillators to construct normal ordered pairs,
and thus did not produce a two-body
part. Furthermore, not even the forward scattering
processes contributed to the effective hamiltonian (see Eq. (7) of
Ref.~\cite{lupe}). However, as we mentioned above, it precisely
due to these two $J_z$-dependent terms that our result for
the Fermi velocity $v$ is consistent with the exact
value to first order in $J_z$.

The spectrum of ${\cal H}_{(1/N)}$ in \eq{anhb} follows directly
from the representation theory of the $c=1$ conformal algebra.
Let $|\Delta N,\Delta D\rangle_W$ be a highest weight state
such that
\bea
\w_0^0 ~|\Delta N ; \Delta D \rangle_W  &=&
\left(\frac{\Delta N}{2\,r}+
r\,\Delta D\right)
|\Delta N ; \Delta D \rangle_W \nl
\wb_0^0 ~|\Delta N ; \Delta D \rangle_W  &=&
\left(\frac{\Delta N}{2\,r}-
r\,\Delta D\right)
|\Delta N ; \Delta D \rangle_W~~~,
\label{vdo'}
\ena
with $\Delta N -2\Delta D= 0~~{\rm mod}~2$. Here $r$ is the radius
given by \eq{radius} and the numbers $\Delta N$ and $\Delta D$ have
the same interpretation as in the Calogero-Sutherland case
\footnote{Note however that $\Delta N$, being the increment in the
number of fermions ({\it i.e.} the number of the spin-up particles)
has nothing to do with $N$ which, here, is the number of sites.}.
Then, it is immediate to verify that
\beq
{\cal H}_{(1/N)}
{}~|\Delta N , \Delta D ; \{k_i\},\{{\overline k}_j\} \rangle_W =
{\cal E}_{(1/N)}
{}~|\Delta N , \Delta D ; \{k_i\},\{{\overline k}_j\} \rangle_W
\label{eig'}
\eeq
where
\beq
{\cal E}_{(1/N)}= \frac{2\pi}{N}\,v\,
\Bigg[\left(\frac{\left(\Delta N\right)^2}{4\,r^2}  +
r^2\,\left(\Delta D\right)^2 +
k+{\overline k} \right)\Bigg] ~~~,
\label{eigen'}
\eeq
with $k=\sum\limits_i k_i$ and
${\overline k}=\sum\limits_j {\overline k}_j~$ being
the right and left levels of the descendant state
$|\Delta N , \Delta D ; \{k_i\},\{{\overline k}_j\} \rangle_W$.
The energy eigenvalues (\ref{eig'}) coincide with those obtained
by calculating the finite size corrections from the Bethe Ansatz
to first order in $J_z$, and are clearly
degenerate when $k \geq 2$ or
${\overline k} \geq 2$.

This degeneracy is removed if we also
take into account the higher order terms in the $1/N$ expansion
of the effective hamiltonian (\ref{hamd}). When the magnetic field is  
not zero,
the first subleading correction is given by ${\cal H}_{(2)}$
of \eq{hwh2}.
Using the generalized Sugawara construction, and then introducing the  
new
$\winf$generators through the Bogoliubov transformation,
we obtain after some straightforward
algebra
$$
{\cal H}_{(2)} \ =\ {\cal H}_{(2)}'\ +\ {\cal H}_{(2)}''
$$
where
\bea
{\cal H}_{(2)}' &=&
-\ \frac{1}{2} \left(\sin\frac{\pi\sigma}{2}
+ \frac{J_z}{\pi} \sin\pi\sigma\right)
\left[ \w^2_0 + \wb^2_0 -\frac{1}{12}
\left( \w^0_0 + \wb^0_0 \right)\right]
\nl
&& -\ \frac{J_z}{2\pi}\sin\pi\sigma
\left( \w^0_0 \,\wb^1_0 +
\w^1_0 \,\wb^0_0 \right) ~~~, \label{h2h'}
\ena
and
\bea
{\cal H}_{(2)}'' &=&
-\frac{J_z}{2\pi}\sin\pi\sigma \sum_{\ell \neq 0}
\left( \w^0_\ell\,\wb^1_\ell +
\w^1_\ell \,\wb^0_\ell \right) ~~~.
\label{h2h''}
\ena
Neither ${\cal H}_{(2)}'$ nor ${\cal H}_{(2)}''$
are diagonal in the basis
$|\Delta N , \Delta D ; \{k_i\},\{{\overline k}_j\} \rangle_W$
previously considered. However, like in the Calogero-Sutherland
model, we can find suitable combinations
of these states that diagonalize simultaneously
${\cal H}_{(2)}'$ and ${\cal H}_{(1/N)}$.
In fact, these two operators commute with each other,
since they are combinations of the generators of the Cartan  
subalgebra
of $\winf$. Notice that ${\cal H}_{(2)}'$ is not
factorized, but
contains a left-right mixing term. This is a new feature that
distinguishes between
the Heisenberg and the Calogero-Sutherland models.
However, as we shall see in a moment, this mixing does not cause
any problem.
Let us denote by ${\cal W}(\Delta N,\Delta D)$ the set of
states that are simultaneously eigenstates of
${\cal H}_{(2)}'$ and ${\cal H}_{(1/N)}$.
It is easy to check that ${\cal H}_{(2)}''$
given in \eq{h2h''} has vanishing expectation value on
any state of ${\cal W}(\Delta N,\Delta D)$. Therefore, according to
ordinary perturbation theory, ${\cal H}_{(2)}''$ gives no  
contribution
to first order in $J_z$, and can be dropped.
The effective hamiltonian of the Heisenberg model up to $O(1/N^2)$
is, then,
\bea
{\cal H}_H &\equiv& {\cal H}_{(1/N)} + \left(\frac{2\pi}{N}\right)^2
{\cal H}_{(2)}'
\nl
&=&\frac{2\pi}{N} \left[ \cos\frac{\pi\sigma}{2}
+ \frac{J_z}{\pi} \left( \cos\pi\sigma + 1\right)\right]
\left( W^1_0\ +\ \wb^1_0 \right)\nl
&&-\ \left(\frac{2\pi}{N}\right)^2\left\{
\frac{1}{2} \left(\sin\frac{\pi\sigma}{2}
+ \frac{J_z}{\pi} \sin\pi\sigma\right)
\left[ \w^2_0 + \wb^2_0 -\frac{1}{12}
\left( \w^0_0 + \wb^0_0 \right)\right]
\right.\nl
&&\left.+\ \frac{J_z}{2\pi}\sin\pi\sigma
\left( \w^0_0 \,\wb^1_0 +
\w^1_0 \,\wb^0_0 \right)\right\}~~~.
\label{hhf}
\ena
and its spectrum can be easily found.
Actually, one can verify that the eigenstates of ${\cal H}_{H}$
have the same form as those of the Calogero-Sutherland model
with $\xi=1$ (see for example Eqs. (\ref{1,0})-(\ref{3,0**})).
We denote these states simply by
$|\,\{n_j\},\{{\overline n}_j\}\,\rangle$,
where $\{n_j\}$ and $\{{\overline n}_j\}$ are the numbers that  
specify
the levels of the right and left particle-hole pairs created on the
highest weight state labeled by $\Delta N$ and $\Delta D$.
Then, a direct calculation shows that
\bea
\w_0^0\,|\,\{n_j\},\{{\overline n}_j\}\,\rangle &=&
Q \ |\,\{n_j\},\{{\overline n}_j\}\,\rangle
 ~~~,\nl
\w_0^1\,|\,\{n_j\},\{{\overline n}_j\}\,\rangle &=&
\left(\frac{1}{2}Q^2 + \sum_{j=1} n_j \right) \,
|\,\{n_j\},\{{\overline n}_j\}\,\rangle ~~~,
\label{w20} \\
\w_0^2\,|\,\{n_j\},\{{\overline n}_j\}\,\rangle &=&
\left(\frac{1}{3}Q^3 + 2Q \sum_{j=1} n_j + \sum_{j=1} n_j^2 -
\sum_{j=1} \left(2j-1\right) n_j \right) \,
|\,\{n_j\},\{{\overline n}_j\}\,\rangle
 ~~~,
\nonumber
\ena
with $Q$ being the right charge given by \eq{vdo'}.
Obviously the states
$|\,\{n_j\},\{{\overline n}_j\}\,\rangle$ are
eigenstates of $\wb_0^0$, $\wb_0^1$ and $\wb_0^2$ also, and their
eigenvalues are given by \eq{w20} with $Q$ and $\{n_j\}$ replaced by
${\overline Q}$ and $\{{\overline n}_j\}$, respectively.
Using this result in \eq{hhf},
it is trivial to compute the
energy of the low-lying excitations up to order $1/N^2$.
In this case, however, the comparison with the exact results from the
Bethe Ansatz is not immediate because only the leading finite size
corrections to the energy are currently available in the literature.

We conclude by pointing out that in the absence of magnetic field,
${\cal H}_{(2)}$ is zero and the
first non vanishing corrections to the conformal hamiltonian are
of order $1/N^3$. If $B=0$, these are
\barr
{\cal H}_{(3)} &=&-\ \frac{1}{6} \left(1+\frac{J_z}{\pi}\right)
\left(\v^3_0 +  \vb^3_0\right)
+ \frac{1}{10} \left(\frac{7}{12}+\frac{J_z}{\pi} \right)
\left(\v^1_0 + \vb^1_0\right) \nl
&&-\ \frac{J_z}{4\pi} \sum_{\ell=-\infty}^{\infty}
\left[\, {\ell}^2 \left( \v^0_{-\ell}\,\v^0_{\ell}
+ \vb^0_{-\ell} \,\vb^0_{\ell}\right) +
\left(2{\ell}^2 +\frac{1}{3}({\ell}^2-1) \right)
\v^0_{\ell} \,\vb^0_{\ell}\right] \nl
&&-\ \frac{J_z}{2\pi}
\sum_{\ell=-\infty}^{\infty}\left(\v^2_{\ell}\, \vb^0_{\ell}
+2\, \v^1_{\ell} \,\vb^1_{\ell} + \v^0_{\ell} \,\vb^2_{\ell}
\right)~~~.
\earr
After performing the Bogoliubov transformation, ${\cal H}_{(3)}$  
becomes
$$
{\cal H}_{(3)} = {\cal H}'_{(3)} + {\cal H}''_{(3)}~~~,
$$
with
\bea
{\cal H}'_{(3)}
&=&-\ \frac{1}{6} \left(1+\frac{J_z}{\pi}\right)
\left(\w^3_0 +  \wb^3_0\right)
+ \frac{1}{10} \left(\frac{7}{12}+\frac{J_z}{\pi} \right)
\left(\w^1_0+ \wb^1_0\right) \nl
&&-\ \frac{J_z}{4\pi} \sum_{\ell=-\infty}^{\infty}
{\ell}^2 \left( \w^0_{-\ell}\,\w^0_{\ell}
+ \wb^0_{-\ell} \,\wb^0_{\ell}\right)
-\frac{J_z}{\pi}\,\w^1_0 \,\wb^1_0~~~,
\label{or31}
\ena
and
\beq
{\cal H}''_{(3)} =
-\ \frac{J_z}{4\pi} \sum_{\ell\not=0} \left[
\left(2{\ell}^2 +\frac{1}{3} {\ell}^2 \right)
\w^0_{\ell} \,\wb^0_{\ell} +4\,\w^1_{\ell}\,\wb^1_{\ell}\right] ~~~.
\label{or3}
\eeq
The procedure is now identical to that previously discussed.
In fact, one can easily realize that ${\cal H}'_{(3)}$ in \eq{or31}
and ${\cal H}_{(1/N)}$ in \eq{anhb} with $\sigma=0$
commute with each other, so that they can be
diagonalized simultaneously. If we take any common eigenstate of  
these
operators, then we can check that ${\cal H}''_{(3)}$ in \eq{or3} has  
always
vanishing expectation value on it. Thus, according to
ordinary perturbation theory,
${\cal H}''_{(3)}$ can be dropped to first order in $J_z$.
Therefore, the effective
hamiltonian of the Heisenberg model with no magnetic field up to
order $1/N^3$ is ${\cal H}_{(1/N)} + (2\pi /N)^3\,{\cal H}'_{(3)}$.

\vskip 1.5cm
\section{Conclusions}
\label{concl}
\bigskip

We conclude by commenting on some of the most relevant and
general features of our method.
First of all, we would like to stress
that our algebraic bosonization can be applied to any (abelian)
gapless fermionic hamiltonian consisting of a bilinear kinetic term
and an arbitrary interaction. No special requirements
on the form of the dispersion relation and the potential
are needed. In particular, it is not necessary for the
system to be integrable.
In lattice models, one limitation is that
no Umklapp terms should appear in the low-energy hamiltonian.
Indeed, these would spoil the
charge-current conservation, which is the origin of the
$\winf$algebraic structure of the effective theory.

Since the Fermi surface is identified from the bilinear part of
the hamiltonian, our procedure is strictly perturbative in
the coupling constant $g$ of the interaction term. Limiting
our analysis to the conformal leading order in the
$1/N$ expansion, it is possible to diagonalize non-perturbatively
the hamiltonian by means of a Bogoliubov
transformation.
However, once we also include the subleading $O(1/N^2)$-part of the
effective hamiltonian, only the first perturbative order
in the coupling constant is meaningful. In fact,
when one also takes into account the $O(1/N^2)$-terms of the
dispersion curve around the Fermi surface, spurious states are
effectively introduced. These would contribute beyond
the first perturbative order, spoiling the finiteness of the theory.
In some cases, however, non-perturbative improvements are possible.
For example, by exploiting some results of the Bethe Ansatz solution
of the Calogero-Sutherland model, we have been able
to write the {\it complete} effective hamiltonian for ${\it any}$  
value
of the coupling constant (see \eq{hcsf1}), and compute its low-energy
spectrum using purely algebraic methods. However,
even if such improvements are possible, for theories with a  
non-trivial
phase diagram we can only hope to reach the phase continously
connected to $g=0$.

Finally, we point out that the complete effective hamiltonian does
not show, in general, a factorization between the left and right  
sectors,
contrarily what happens at the leading conformal order (see
\eq{confo}). However, the left-right mixing can
only occur
through the zero modes of the generators $\winf \times {\overline
\winf}$algebra (see for example the hamiltonian of the Heisenberg  
model
given in \eq{hhf}). Hence, it is still possible to use
the representation theory of the chiral $\winf$algebra to compute the
low-energy spectrum of the model.

\vskip 2.5cm
\noindent
{\large {\bf {Acknowledgements}}}
\vskip 0.5cm
\noindent
We would like to thank the organizers of the Benasque Center for  
Physics
for their kind hospitality during the early stages of this work. \\
G.R.Z. thanks the Dipartimento
di Fisica Teorica dell' Universit\'a di Torino and I.N.F.N. (sezione  
di
Torino) for support and hospitality during the period in which
most part of this work has been done.\\
This research was partially supported by MURST and the EU,
within the framework of
the program ``Gauge Theories, Applied Supersymmetry
and Quantum Gravity'',
under contract SCI*-CT92-0789.
\vfill
\break

\appendix
\section{\bf The mathematics of the $\winf$algebra}
\bigskip

In this appendix we collect several mathematical results concerning
the $\winf$algebra, in particular we give the complete expression of  
the
algebra in a compact form, a survey of representation theory and
briefly discuss the characters of the representations.
All of these results are taken from the original references
\cite{kac1,kac2,ctz5}.
We also quote here some formulae regarding the expression of the
$\winf$generators on the conformal plane and on the cylinder.

\bigskip
\noindent{\bf The complete form of the $\winf$algebra}

The complete $\winf$algebra is expressed in compact
form by using a parametric sum of the $\v^i_n$ current modes, denoted
by
$V\left(- z^n \exp (\lambda D)\right)$, where
$D\equiv z{\partial\over \partial z}$ \cite{kac1}.
These satisfy the algebra,
\barr
{[\ V\left(-z^r {\rm e}^{\lambda D} \right)\ ,\
V\left(-z^s {\rm e}^{\mu D} \right)\ ]} &= &
\left( {\rm e}^{\mu r} -{\rm e}^{\lambda s} \right)
\ V\left(- z^{r+s}\ {\rm e}^{(\lambda +\mu) D} \right)\ \nonumber\\
&\ & +\ c\ \delta_{r+s,0}\  {{\rm e}^{-\lambda r} -{\rm e}^{-\mu s}
\over
 1 - {\rm e}^{\lambda +\mu} } \ ,
\label{wconp}\earr
where $c$ is the central extension.
The currents $\v^i_n$ of conformal spin $h=i +1 \ge 1$ and
mode index $n\in{\bf Z}$, are identified by expanding this
parametric operator in $\lambda$, namely
\beq
V^i_n \equiv V \left(- z^n f^i_n (D) \right) \ ,
\label{wexp}\eeq
where $f^i_n (D)$ are specific $i$-th order polynomials which
diagonalize the central term of \eq{wconp} in the $i,j$ indices
\cite{kac1,ctz4}. For example, we have
\barr
\v^0_n &\equiv & V\left(-z^n\right)\ ,\nonumber\\
\v^1_n &\equiv & V\left( -z^n \left(D + {n+1\over 2}\
\right)\right)\ ,\nonumber\\
\v^2_n &\equiv & V\left( -z^n \left(D^2+(n+1)D
+ {(n+1)(n+2)\over 6}\right)\right)\ ,\nonumber\\
\v^3_n &\equiv & V\left( -z^n \left(D^3+{3\over
2}(n+1)D^2+{\frac{(6n^2+15n+11)}
{10} }D
\right. \right.\nonumber\\
&&\qquad\qquad\qquad\left. \left.
+ {\frac{(n+1)(n+2)(n+3)}{20} }\right)\right)\ .
\label{vex}
\earr

\bigskip
\noindent{\bf A survey on representation theory}

All unitary irreducible quasi-finite highest-weight
representations
\cite{kac2}, denoted by $M\left(\winf,c,\vec{Q}\right)$, exist when
the central charge $c=m$ is a positive integer,
and are characterized by a highest weight state
$|\,\vec{Q}\,\rangle_W$, which satisfies
\beq
V\left(- z^n {\rm e}^{\lambda D}\right) \vert \,\vec{Q}\,\rangle_W =
0~~~,
\quad n>0~~~,
\label{whwcc}
\eeq
and
\beq
V\left( - {\rm e}^{\lambda D}\right) \vert\, \vec{Q}\,\rangle_W =
\Delta(\lambda)\vert \,\vec{Q}\,\rangle_W \equiv
\sum_{i=1}^m {{\rm e}^{\lambda Q_i} -1 \over {\rm e}^\lambda -1 } \
\vert \,\vec{Q}\,\rangle_W ~~~,
\label{weigen}
\eeq
where $\vec{Q}=\{Q_1,\dots,Q_m\}\in {\bf R}^m$.
In particular, the eigenvalues of the operators $\v^0_0$ and $\v^1_0$
given in \eq{wcs} can be recovered by expanding $\Delta(\lambda)$
and comparing to \eq{vex}.
The infinite tower of states (Verma module) in each
representation is generated by expanding in the $\{\lambda_i\}$ of
\beq
V\left(- z^{-n_1} {\rm e}^{\lambda_1 D} \right) \cdots
V\left(- z^{-n_k} {\rm e}^{\lambda_k D} \right) \vert \vec{Q}\
\rangle_W~~~,
\qquad n_1 \ge n_2 \cdots \ge n_k >0 ~~~,
\label{wverma}\eeq
where $n=\sum\limits_{i=1}^k n_i $ is the level of the states.
The quasi-finite representations have only a finite number of
independent states at each level,
thus there are an infinity of polynomial relations
among the generators $\v^i_n$, whose explicit form depends on the
values of $c$ and $\vec{Q}$.
The number of independent states $d(n)$ at level $n$ is encoded in
the (specialized) character of the
representation ($\vert q\vert<1 $) \cite{bpz},
\beq
\chi_{M(\winf,m,\vec{Q})} (q)
\equiv {\rm tr}_{M(\winf,m,\vec{Q})}
\left( q^{\scr{\v^1_0-{m\over 24}}} \right) =
q^{\scr{\sum\limits_{i=1}^m \left({Q_i^2\over 2} -{1\over 24}\right)}  
}
 \sum_{n=0}^\infty d(n) q^n~~~.
\label{wchar}\eeq
A representation is called {\it generic} if the weight $\vec{Q}$
has components $(Q_i -Q_j ) \not\in {\bf Z}\ , \ \forall i\neq j$,
and {\it degenerate} if it has $(Q_i-Q_j) \in {\bf Z}$ for some
$i\neq j$.
The weight components $\{Q_i\}$ of the degenerate representations
can be grouped and ordered in congruence classes modulo ${\bf
Z}$
\cite{kac2},
\barr
&& \{ Q_i,\dots,Q_m \}=\{ s_1+n^{(1)}_1,\dots,s_1+n^{(1)}_{m_1} \}
\cup \cdots
\cup \{ s_k+n^{(k)}_1,\dots,s_k+n^{(k)}_{m_k} \}\ ,\nonumber\\
&& n^{(i)}_j \in {\bf Z}\ ,\ n^{(i)}_1 \ge n^{(i)}_2 \ge\cdots
\ge n^{(i)}_{m_i} ~~,~~ m=\sum_{i=1}^k m_i ~~,~~s_i \in {\bf R}~~~.
\label{congu}\earr
A two-class representation is the tensor product of two
one-class representations. Therefore, the one-class degenerate
representations are the basic building blocks, which one can use
to construct the $\winf$minimal models \cite{ctz5}.
The character for the generic representations is
\beq
\chi_{M(\winf,m,\vec{Q})} (q) = \prod_{i=1}^m \
{q^{Q_i^2/2}\over\eta(q)} =
\prod_{i=1}^m \ \chi_{M(\widehat{U(1)},1,Q_i)} (q) ~~~,
\label{chir}\eeq
where $\eta(q)$ is the Dedekind function,
\beq
\eta(q)=q^{1/24} \prod_{n=1}^\infty \left(1-q^n\right)~~~.
\label{dede}\eeq
\eq{chir} exhibits the form of the $\winf$character
in terms of $m$ characters of the $\widehat{U(1)}$ algebra,
which implies a one-to-one equivalence of generic $\winf$and
$\widehat{U(1)}^{\otimes m}$ representations \cite{kac2,ctz5}.
The character for the one-class degenerate representations is
\barr
\chi_{M(\winf,m,\vec{Q})} (q) &=& \eta(q)^{-m}\  
q^{\sum\limits_{i=1}^m
Q_i^2/2}
\prod_{1\le i<j\le m} \left( 1-q^{n_i-n_j+j-i} \right)~~~,\nonumber\\
\vec{Q} = \{Q_1,\dots,Q_m\} &=& \{s+n_1,\dots,s+n_m\}~~,\quad
n_1\ge\cdots \ge n_m~~~.
\label{chid}\earr
Note that the number of independent states $d(n)$ at level $n$ is
smaller for degenerate (\ref{chid}) than for generic (\ref{wchar})
representations, because the former have additional relations
among the states that lead to null vectors.
This is the origin of reducibility of the $\widehat{U(1)}^{\otimes  
m}$
representations with respect to the $\winf$algebra.

In order to construct a $\winf$theory, one should combine $\winf$
representations that are closed under the fusion rules \cite{bpz}.
For {\it generic} $\winf$\reps, the fusion rules require that all
highest weight vectors $\vec{Q}$ span a lattice $\Gamma$,
\beq
\Gamma=\left\{  \vec{Q}\ \Big\vert\ \vec{Q}=
\sum_{i=1}^m n_i \vec{{\bf v}}_i\ ,\quad n_i \in {\bf Z} \right\}  
~~~,
\label{latt}
\eeq
whose points satisfy $(Q_i-Q_j)\not\in {\bf Z},\ \forall\ i\neq j$
(see \cite{ctz5} for more details).
The resulting $\winf$theory can be associated to a system
with $m$ components at the Fermi point with the basis vector
$\vec{{\bf v}}_i$ representing a physical elementary excitation
in the $i$-th component of the Fermi point.

\bigskip
\noindent{\bf The $\winf$operators on the cylinder}

It is well-know that the form
(\ref{wcurp}) of the $\winf$currents on the plane,
constructed out of the operators (\ref{weypl}),
is different from that of the physical currents on the cylinder,
constructed out the operators (\ref{f+}) and (\ref{f-}).
This is due to a normal ordering effect (for a review
see Ref.~\cite{car}).
To obtain their specific form on the cylinder in the fermionic case
($c=1$), one applies the conformal mapping (\ref{cft}) to each  
fermion
field in (\ref{wcurp}), paying attention to the different normal  
ordering
in the plane and the cylinder (for a detailed account see  
\cite{ctz4}).
For example,
\barr
V^0_R (u)&=&:F^\dagger (u)\,F (u):~\equiv\lim_{u_1,u_2\to u}
         \left(F^\dagger (u_1)\,F (u_2) - {1\over u_1-u_2} \right)
\nonumber\\
      &=& {dz\over du}\ V^0(z) \ + \lim_{u_1,u_2\to u}\left[
        \left({dz_1\over du_1}{dz_2\over du_2}\right)^{1/2}
         {1\over z_1-z_2} -{1\over u_1-u_2} \right]\nonumber\\
      &=&  {z\over R}\ V^0(z) \ .
\label{winfcyl}\earr
Proceeding similarly for the other currents, one can obtain
the explicit relations between the zero modes
in the two geometries; for the first
few values of the spin these are
\barr
\left(V_R\right)^0_0 &=& \v^0_0~~~,\nonumber\\
\left(V_R\right)^1_0 &=& \frac{1}{R}\left(\,V^1_0-
\frac{1}{24}\, \right)~~~,\nonumber\\
\left(V_R\right)^2_0 &=& \frac{1}{R^2}\left(\,\v^2_0-
\frac{1}{12}\,\v^0_0\,\right)~~~,\nonumber\\
\left(V_R\right)^3_0 &=& \frac{1}{R^3} \left(\, \v^3_0-
\frac{7}{20}\, V^1_0 - \frac{7}{960}\, \right)~~~.
\label{wopcyl}\earr

\end{document}